\documentclass[preprint,12pt]{elsarticle}

\usepackage{amssymb,amsmath}
\usepackage{latexsym}
\usepackage{amsthm}
\usepackage{graphicx}
\usepackage{geometry}                % See geometry.pdf to learn the layout options. There are lots.
\usepackage{color}
\usepackage{xcolor}
\usepackage{pdfsync}
\usepackage{fancyhdr}
\usepackage[applemac]{inputenc}
\usepackage{amsfonts} % needed for bold Greek, Fraktur, and blackboard bold
\usepackage{graphicx} % needed for figures
\usepackage{caption}
\usepackage{subcaption}

\usepackage[breaklinks,colorlinks,backref]{hyperref}
\hypersetup{
    colorlinks, %set true if you want colored links
    linktoc=all, %set to all if you want both sections and subsections linked
    linkcolor=red,  %choose some color if you want links to stand out
  citecolor=hyptxt,
  urlcolor=blue}
  \hypersetup{
  citebordercolor=,
  filebordercolor=green,
  linkbordercolor=blue
}
\definecolor{hyptxt}{rgb}{0.7, 0.4, 0.9}
\usepackage{bibtopic}

\catcode `\@=11
\@addtoreset{equation}{section}

\font\teneufm=eufm7 at 11pt
\font\seveneufm=eufm7 at 8pt
\font\fiveeufm=eufm7 at 5pt
\newfam\eufmfam
\textfont\eufmfam=\teneufm
\scriptfont\eufmfam=\seveneufm
\scriptscriptfont\eufmfam=\fiveeufm

\def\R{\relax\ifmmode {{\rm I} \! {\rm R}}\else${{\rm I} \! {\rm R}}$\fi}
\def\R{\relax\ifmmode {{\rm I} \! {\rm R}}\else${{\rm I} \! {\rm R}}$\fi}
\def\Z{\relax\ifmmode {\Bbb Z}  \else${\Bbb Z}$\fi}
\def\C{\relax\ifmmode {\Bbb C}  \else${\Bbb C}$\fi}
\def\F{\relax\ifmmode {\Bbb F}  \else${\Bbb F}$\fi}
\def\N{\relax\ifmmode {\Bbb N} \else${\Bbb N}$\fi}
\def\H{\relax\ifmmode {\cal H} \else${\cal H}$\fi}
\catcode `\@=12

\newcommand{\be}{\begin{equation}}
\newcommand{\ee}{\end{equation}}
\newcommand{\en}{\end{equation}}

\newcommand{\ba}{\begin{eqnarray}}
\newcommand{\ea}{\end{eqnarray}}

\newcommand{\beano}{\begin{eqnarray*}}
\newcommand{\enano}{\end{eqnarray*}}

\newcommand{\bei}{\begin{itemize}}
\newcommand{\eni}{\end{itemize}}

\newcommand{\bee}{\begin{enumerate}}
\newcommand{\ene}{\end{enumerate}}

\newtheorem{prop}{Proposition}[section]

\newcommand{\beprop}{\begin{prop}}
\newcommand{\enprop}{\end{prop}}
\newcommand{\bprf}{\begin{proof}}
\newcommand{\eprf}{\end{proof}}

\newcommand{\nn}{\nonumber}

\newcommand{\ud}{\mathrm{d}}
\def\rg{\rangle}
\def\lg{\langle}

\def\eal{e^{(\alpha)}}

\def\sfL{\mathsf{L}}
\def\sfH{\mathsf{H}}
\def\LL{L^2(\R_+^\ast, \ud x)}
\def\ii{\mathrm{i}}

\journal{Annals of Physics}

\begin{document}

\begin{frontmatter}

%% Title, authors and addresses

%% use the tnoteref command within \title for footnotes;
%% use the tnotetext command for theassociated footnote;
%% use the fnref command within \author or \address for footnotes;
%% use the fntext command for theassociated footnote;
%% use the corref command within \author for corresponding author footnotes;
%% use the cortext command for theassociated footnote;
%% use the ead command for the email address,
%% and the form \ead[url] for the home page:
%% \title{Title\tnoteref{label1}}
%% \tnotetext[label1]{}
%% \author{Name\corref{cor1}\fnref{label2}}
%% \ead{email address}
%% \ead[url]{home page}
%% \fntext[label2]{}
%% \cortext[cor1]{}
%% \address{Address\fnref{label3}}
%% \fntext[label3]{}

\title{Three examples of quantum dynamics on the half-line with smooth bouncing}

%% use optional labels to link authors explicitly to addresses:
%% \author[label1,label2]{}
%% \address[label1]{}
%% \address[label2]{}

\author{C. R. Almeida}

\address{Universidade Federal do Esp\'irito Santo, Vit\'oria,  29075-910, ES, Brazil}
\ead{carlagbjj@hotmail.com} 

\author{H. Bergeron}
\address{ISMO, UMR 8214 CNRS, Univ Paris-Sud,  France}
\ead{herve.bergeron@u-psud.fr}

\author{J.P. Gazeau}

\address{Laboratoire APC, UMR 7164, Univ Paris  Diderot, Sorbonne Paris-Cit\'e 75205 Paris, France}
\ead{gazeau@apc.univ-paris7.fr}

\author{A.C. Scardua}
\address{Centro Brasileiro de Pesquisas F\'isicas,
	Rua Xavier Sigaud 150, 22290-180 - Rio de Janeiro, RJ, Brazil}
\ead{ arthur@scardua.me}

\begin{abstract}
This article is an introductory presentation of the  quantization of the half-plane based on  affine coherent states (ACS). The half-plane is viewed as the phase space for the dynamics of a positive physical quantity evolving with time, and its affine symmetry is preserved due to the covariance of this type of quantization.  We promote the interest of such a procedure for transforming a classical model into a quantum one, since the singularity at the origin is systematically removed, and the arbitrariness of boundary conditions can be easily overcome. We explain some  important mathematical aspects of  the method. Three elementary examples of applications are presented,  the quantum breathing of a massive sphere, the quantum smooth bouncing of a charged sphere, and a smooth bouncing of ``dust'' sphere as a simple model of quantum Newtonian cosmology. 
\end{abstract}

\begin{keyword}
%% keywords here, in the form: keyword \sep keyword
Integral quantization \sep Half-plane \sep Affine symmetry \sep Coherent states \sep Quantum smooth bouncing 
%% PACS codes here, in the form: \PACS code \sep code

%% MSC codes here, in the form: \MSC code \sep code
%% or \MSC[2008] code \sep code (2000 is the default)

\end{keyword}

\end{frontmatter}

%% \linenumbers

%% main text

%\documentclass[11pt]{article}

%%%%%%%%%%%%%%%%%%%%%%%%%%%%%%%%%%%%%%%%%%%%%%%%%%%%%%%%%%%%%%%

\date{\today}

%\maketitle

\tableofcontents

\section {Introduction}
\label{intro}

The basic, or so-called canonical, quantization procedure \cite{dirac82} for the motion of a particle on the line
consists in transforming  pairs of canonical variables $(q,p)$ in the corresponding phase space  $\R^2$ into a non-commuting pair of self-adjoint operators in some Hilbert space, e.g. the space of square integrable complex-valued functions on the line, 
\begin{equation}
 (q,p)  \mapsto  (Q,P)\,; \quad [Q,P] = \ii \hbar I\, ; \ Q\psi(x)= x\psi(x)\, ; \ P\psi(x)= -\ii\hbar \frac{\partial}{\partial x}\psi(x)\, . 
\end{equation}
The procedure is extended to the quantization of \textit{classical observables} $f(q,p)$,
\begin{equation}
f(q,p) \mapsto f(Q,P) \mapsto (\mathrm{Sym}f)(Q,P)\, ,
\end{equation}
where $\mathrm{Sym}$ stands for a certain choice of symmetrisation  of the operator-valued function $f(Q,P)$. Besides the above ordering ambiguity, the procedure immediately raises deep questions about its domain of validity. What about singular $f$, e.g. the phase or angle function $\arctan(p/q)$? What about other phase space geometries which are limited by impassable boundaries? Despite their elementary aspects, these singular geometries leave open many questions both on mathematical and physical levels, irrespective of the variety of quantization methods \cite{alienglis05,zachos05,curtright16}. Indeed, most of  the latter, despite their aesthetic mathematical content, are too demanding for models to be quantized. 

This article is precisely devoted to one of the most elementary examples of such geometries, namely the half-plane  $\{(q,p)\,  | \,  q>0\,,\, p\in \R\}$ corresponding to a motion on the positive half-line, the origin $x=0$ being viewed as an inaccessible singularity.  It  is deemed a pedagogical introduction to affine covariant integral quantization of functions on the half-plane and its applications. This procedure  has been introduced recently \cite{bergaz14,berdagama14} for providing smooth solutions to singularity problems in early quantum cosmology (see \cite{beczgama16} and most recent references therein). It is consistent with the phase space symmetries of the system, and carries the name of the group that represents such symmetries.

 Let us explain more about the term ``\textit{affine covariant integral quantization}''.  The  adjective ``affine'' refers to the group of symmetries of the half-plane combining translations and  dilations. This symmetry is different from the translational symmetry of the plane on which is based the familiar canonical or Weyl-Heisenberg quantization \cite{bergaz14}. ``Covariant'' means that the quantization map intertwines classical (geometric operation) and quantum (unitary transformations) symmetries. Integral means that we use all ressources of integral calculus, in order to implement the method when we apply it to  singular functions, or distributions, for which the integral calculus is an essential ingredient. 

Classical physics is rich in one-dimensional models with law determining the time behavior of a positive dynamical physical quantity, like the position $x(t)$ of a particle moving on the positive half-line $\R_+^\ast= \{x\in \R\,  | \,  x>0\}$, its kinetic energy, a length $l(t)$, the radius $r(t)$ of a sphere, and many other examples, like in optomechanics the distance between a fixed mirror and a moveable mirror,  a small vibrating element that forms one of the end mirrors of a Fabry-Perot cavity \cite{aspelmeyer14}, or like  those involving the dynamics of the Hubble scale factor in Cosmology (see Chapter 1 in \cite{mukhanov05}). In each of these cases, the origin or the value   $x=0$ is considered as a classically impenetrable barrier. Due to the restriction $x>0$, this barrier is more than a simple singularity. It is purely geometrical and it should not be confused with a Dirac potential at the origin, $V(x) = k\delta(x)$, or a singular potential like $V(x) = k/x$ (Kepler-Coulomb on the half-line), or others, for which the position $x=0$ is supposed to be accessible. On the classical level, such a geometric singularity $x=0$ is, in principle, attainable at the price to deal with infinite quantities, like an infinite acceleration in the case of the reflection of the particle. In each case of such dynamical models, the corresponding phase space, i.e. the set of initial positions and momenta for any motion on the half-line,  is the positive half-plane $\R_+^\ast\times \R=\{(q,p)\,  | \,  q>0\,,\, p\in \R\}$. This geometry has a nice group structure, and this will be the rationale backing our quantization method based on affine coherent states (ACS) or, equivalently, wavelets \cite{aagbook14}.  ACS quantization is a particular approach pertaining to covariant affine integral quantization \cite{gazmur16} (see also the recent extension to the motion in the punctured plane \cite{gazkoimur17}). 

The organisation of the paper is as follows. In Section \ref{affG} we present the geometry of the half-plane and its particular symmetry which underlies a group structure, namely the affine group of transformations ``$ax +b$'' of the real line. This group has two unitary irreducible representations only, and we use one of them to build our affine coherent states, similarly to the continuous wavelet construction in signal analysis \cite{aagbook14}.  Section \ref{ACSquant} is devoted to the ACS quantization  with  its principal definition and implementation formulas in Subsection \ref{ACSquant1}, while the subsequent ACS mean values or \textit{semi-classical portraits} of operators are presented in Subsection \ref{semclassport}. In order to illustrate our method with simple models, we give in Section \ref{lagham} a brief survey of Lagrangian and Hamiltonian mechanics appropriate to the motion of the half-line and its phase space, the necessary formalism for implementing ACS quantization of Hamiltonian dynamics in Section \ref{acshalfline}. Three simple and illuminating examples are then presented, in Section \ref{msphere} with a ``breathing'' massive  sphere,  in Section \ref{chsphere} with a ``bouncing'' charged sphere, and in Section \ref{dust} with a bouncing ``dust'' sphere as an elementary model of Newtonian cosmology, the latter one being the most developed in terms of dynamical evolution.  Our results are discussed in Section \ref{conclus}, where we also list some future perspectives.

Since this article is intended   to be a pedagogical initiation to the ACS quantization, we start with the basics of the procedure, progressively evolving into the applications in form of examples cited above. As expected, the quantum correction in the semi-classical approach eliminates the singularity at the origin, creating the bounce mentioned above. Appendices are devoted to the most elaborate part of the mathematical formalism. Thus, the reader is expected to go through the main text without serious difficulty, and to revisit the appendices if there are any doubts about the mathematical background.

\section{The affine group and its representation $U$} 
\label{affG}

The half-plane can be viewed as the phase space for the (time) evolution of a positive physical quantity, for instance  the position of a particle moving in the half-line. Let  the upper half-plane $\Pi_{+}:=\{(q,p)\,|\, q>0\,,\,p\in\mathbb{R} \} \simeq  \mathbb{R}_{+}^{\ast}\times\mathbb{R}$ be equipped with the  uniform measure $\mathrm{d}q\,\mathrm{d}p$. Together
with the multiplication
\begin{equation}
\label{multaff} 
(q,p)(q_{0},p_{0})= \left( qq_{0}, \frac{p_{0}}{q}+p \right),\, q\in\mathbb{R}_{+}^{\ast},\, p\in\mathbb{R}\, ,
\end{equation}
the unity $(1,0)$ and the inverse
\begin{equation}
\label{invaff} 
(q,p)^{-1}= \left(\frac{1}{q}, -qp \right)\, ,
\end{equation}
$\Pi_{+}$ is viewed as the affine group Aff$_{+}(\mathbb{R})$ of the real line, i.e., the two-parameter group of transformations of the line defined by
\begin{equation}
\label{affgrline}
\R \ni x \mapsto (q,p)\cdot x= \frac{x}{q} + p\, . 
\end{equation}
We have chosen the standard (Liouville) phase space measure $\mathrm{d}q\,\mathrm{d}p$ because it is invariant with respect to the  left action of the affine group on itself
\begin{equation}
\label{linvmeasure}
\mathrm{Aff}_{+}(\mathbb{R})\ni (q,p) \mapsto (q_0,p_0)(q,p)= (q^{\prime},p^{\prime})\,, \quad  \mathrm{d}q^{\prime}\,\mathrm{d}p^{\prime}= \mathrm{d}q\,\mathrm{d}p\,. 
\end{equation} 
Note that if we instead consider the right action $(q,p) \mapsto (q,p)(q_0,p_0)$, the corresponding invariant measure would be $\mathrm{d}q\,\mathrm{d}p / q$.

The affine group Aff$_{+}(\mathbb{R})$ has two non-equivalent unitary irreducible representations (UIR) $U_{\pm}$ \cite{gelnai47,aslaklauder68} (see \ref{Grouprep} for a concise explanation about this terminology). Both are square integrable, i.e. $\int_{\Pi_{+}} \mathrm{d}q\,\mathrm{d}p \vert\lg \phi |U_{\pm}(q,p) \phi\rg\vert^2 < \infty$ for all $\phi$ in a dense subset of the Hilbert space carrying the representation $U_{\pm}$, and this is the rationale behind \textit{continuous wavelet analysis} \cite{wave89}. Without loss of generality, only the UIR $U_{+}\equiv U$ is concerned from now on. This representation is realized in the Hilbert space
$L^{2}(\mathbb{R}_{+}^{\ast},\mathrm{d}x)$ as 
\begin{equation}
\label{affrep+}
(U(q,p)\psi)(x)=\frac{e^{\ii px}}{\sqrt{q}}\psi \left( \frac{x}{q} \right)\,.
\end{equation}
The above Hilbert space is actually the Fourier image of functions on the line which can be extended analytically to the upper half-plane (Hardy space \cite{grafakos09}).  

\section{Quantization with affine coherent states (ACS)}
\label{ACSquant}

\subsection{ACS quantization}
\label{ACSquant1}
Let us implement the affine integral covariant  quantization, which is  described in \ref{covintquant} in its generality, by restricting the method to the specific case of rank-one density operator or projector $\rho=|\psi\rangle\left\langle \psi\right|$, where $\psi$ is a unit-norm state and also square integrable on $\mathbb{R}_{+}^{\ast}$ equipped with the measure $\mathrm{d}x/x$ 
\begin{equation}
\label{psicond}
\psi \in L^{2}(\mathbb{R}_{+}^{\ast},\mathrm{d}x) \cap
L^{2}(\mathbb{R}_{+}^{\ast},\mathrm{d}x/x)\, . 
\end{equation}
This $\psi$ is also called ``fiducial vector'' or ``wavelet''. 

Now, we recall and extend a set of results already given in previous works \cite{berdagama14}. The action of the UIR $U$ produces all affine coherent states (ACS), i.e. wavelets, defined as 
\begin{equation}
\label{ACSdef}
|q,p\rangle_{\psi} :=U(q,p)|\psi\rangle \,.
\end{equation}
In the sequel we simplify the notation as $|q,p\rangle_{\psi}= |q,p\rangle$, unless we need to specify the fiducial vector.  The unit norm states \eqref{ACSdef} are not orthogonal. Their overlap is given by the Fourier transform of functions with support on the half-line
\begin{equation}
\label{overlap}
\lg q,p|q^{\prime},p^{\prime}\rg = \int_0^{\infty} \ud x \, e^{\ii (p^{\prime}-p)x} \, \overline{\psi_q(x)}\, \psi_{q^{\prime}}(x)\, ,
\end{equation}
with $\psi_q(x) := \frac{1}{\sqrt{q}}\psi\left(\frac{x}{q}\right)$ is obtained from $\psi$ by unitary dilation. The affine coherent states \eqref{ACSdef} satisfy the resolution of identity in the Hilbert space $L^{2}(\mathbb{R}_{+}^{\ast},\mathrm{d}x)$,
\begin{equation}
\label{affresunit}
\int_{\Pi_{+}}|q,p\rangle\langle q,p|\,\dfrac{\mathrm{d}q\mathrm{d}p}{2\pi c_{-1}}=I\,,
\end{equation}
where
\begin{equation}
\label{cgamma}
c_{\gamma}(\psi):=\int_{0}^{\infty}|\psi(x)|^{2}\,\frac{\mathrm{d}x}{x^{2+\gamma}}\,.
\end{equation}
A detailed proof of the crucial identity \eqref{affresunit} is given in \ref{proofresunit}. Thus, a necessary condition to have \eqref{affresunit} true is that $c_{-1}(\psi) < \infty$, which implies $\psi(0) = 0$, a well-known requirement in wavelet analysis, and which explains the initial request on $\psi$ to be square integrable with respect to $\ud x/x$. Actually \eqref{affresunit}  is the illustration of a general result derived from  the irreducibility and square-integrability of the UIR $U$, and  the application of Schur's Lemma \cite{barracz77} (see \ref{schur}).  

In the sequel we will often simplify the notation as $c_{\gamma}(\psi)= c_{\gamma}$, unless we need to specify the fiducial vector.

The ACS quantization reads as the map that transforms a function (or distribution) on the phase space into an operator in $L^{2}(\mathbb{R}_{+}^{\ast},\mathrm{d}x)$:
\begin{equation}
\label{quantfaff} 
f(q,p)\ \mapsto\ A_{f}=\int_{\Pi_{+}}f(q,p) \,|q,p\rangle\langle
q,p| \, \dfrac{\mathrm{d}q\mathrm{d}p}{2\pi c_{-1}}\,.
\end{equation}
This map is covariant with respect to the unitary affine action $U$:
\begin{equation}
\label{covaff} 
U(q_0,p_0) A_f U^{\dag}(q_0,p_0) = A_{\mathfrak{U}(q_0,p_0)f}\, ,
\end{equation}
with
\begin{equation}
\label{covaff2}
 \left(\mathfrak{U}(q_0,p_0)f\right)(q,p)=
f\left((q_0,p_0)^{-1}(q,p)\right)= f\left(\frac{q}{q_0},q_0(p
-p_0) \right)\, ,
\end{equation}
$\mathfrak{U}$ being the left regular representation of the affine group when $f\in L^2(\Pi_+, \ud q\, \ud p)$. The symmetry property \eqref{covaff} means, and this is certainly the cornerstone of the method, that no point in the phase space $\Pi_{+}$ is privileged. Precisely, the choice of the origin $(1,0)\in \Pi_+$ for the affine geometry of $\Pi_+$ is totally arbitrary, and this is reflected in the unitary map \eqref{covaff}. From now on, our choice of fiducial vector in \eqref{ACSdef} is restricted to real-valued functions, to simplify. Formulas derived with a complex fiducial vector are slightly more involved, but their physical content is not changed. 

One interesting feature of the  map \eqref{quantfaff} lies in the quantization of the phase space point $(q_0,p_0)$ described by the Dirac peak $$\delta(q-q_0)\,\delta(p-p_0)\equiv \delta_{(q_0,p_0)}(q,p)\, .$$ 
Its quantum counterpart is the ACS projector 
\begin{equation}
\label{quantdelta}
\delta_{(q_0,p_0)}(q,p) \mapsto A_{\delta_{(q_0,p_0)}} =  \frac{|q_0,p_0\rangle\langle
q_0,p_0|}{2\pi c_{-1}}\, .
\end{equation}
The deep meaning of this expression will be explained in the part devoted to semi-classical portraits ($\sim$ lower symbols).

Next, we  obtain from the general formulas proven in \ref{proofresunit} the affine quantum versions of the following elementary functions. 
\begin{equation}
\label{quantqp} 
A_{p}=-i\frac{\partial}{\partial x}= P\,,\quad
A_{q^{\beta}}=\frac{c_{\beta-1}}{c_{-1}} \,Q^{\beta}\,,\quad
Qf(x)=xf(x)\,.
\end{equation}
The multiplication operator $Q$ is (essentially) self-adjoint with spectrum equal to the positive half-line. Its spectral decomposition reads as 
\begin{equation}
\label{spdecQ}
Q= \int_{0_+}^{+\infty} \lambda\, \ud E_Q(\lambda)\, , \quad \ud E_Q(\lambda)\equiv |\lambda\rg\lg\lambda| \, \ud \lambda\quad  \lambda>0\, , 
\end{equation}
where $\lg x|\lambda\rg = \delta_\lambda(x)$. On the other hand, the  operator $P$ is symmetric but has no self-adjoint extension \cite{reedsimon75}, as it is expected from the canonical commutation rule, which holds here up to an irrelevant multiplicative constant,
\begin{equation}
\label{comrule}
 [Q,P]= i \frac{c_0}{c_{-1}} I \,.
\end{equation}
 Indeed, we know that the latter holds  true with a pair of self-adjoint operators if both have the whole real line as a spectrum, contrarily to the present case where the operator $Q$ is semi-bounded. If the presence of the constant factor $c_0/c_{-1}$ is considered as a problem, it is always possible to make it equal to 1 through a specific choice of the fiducial vector $\psi$, or by rescaling the ACS as $|q,p\rg \mapsto | \kappa q,p\rg$ with $\kappa = c_0/c_{-1}$, as explained in \ref{proofresunit}. 

The quantization of the product $q\,p$ yields:
\begin{equation}
\label{quantqpdil} 
A_{qp}= \frac{c_0}{c_{-1}}\frac{Q P + PQ}{2} = \frac{c_0}{c_{-1}}\, D\,,
\end{equation}
where $D$ is the dilation generator.  As one of the two generators (with $Q$)  of the UIR $U$ of the affine group, it is essentially self-adjoint. The quantization of the kinetic energy  (up to a factor) of the  particle gives
\begin{equation}
\label{qkinener} 
A_{p^{2}}=P^{2}+\frac{K_{\psi}}{Q^2}\,,\quad K_{\psi}:=\int_{0}^{\infty}
(\psi^{\prime}(x))^{2}\,x\frac{\mathrm{d}x}{c_{-1}}>0\,.
\end{equation}
Therefore, ACS quantization prevents a quantum free particle moving on the positive line from reaching the origin. It is well known that the operator $P^{2}=-\ud^{2}/\ud x^{2}$ in $L^{2}(\R_{+}^{\ast},\mathrm{d}x)$ is not essentially self-adjoint, whereas the above regularized operator, defined on the domain of
smooth function of compact support, is essentially self-adjoint for $K_{\psi}\geq3/4$ \cite{reedsimon75}. Thus, quantum dynamics of the free motion is unique with a suitable choice of the fiducial vector or of the rescaling the parameter $q$ of the wavelet. We should insist with Reed and Simon in \cite{reedsimon75}, p. 145, that the existence of a continuous set of self-adjoint extensions for the operator $P^2$ alone corresponds to the existence of \emph{different physics} for this problem. They are distinguished by boundary conditions at the origin, $\psi^{\prime}(0) + a\, \psi(0) = 0$ for finite real $a$, and $ \psi(0) = 0$ for $a=\infty$,  which are imposed on functions $\psi$ in their respective extension domains. The physical interpretation of these conditions lies in a dependent change of phase for functions behaving like incoming and outgoing  plane waves near the origin where they reflect. No such ambiguity exists with our approach as soon as the factor $K_{\psi}$ is adjusted to a value $\geq3/4$. Moreover, as we illustrate below with our examples, the reflection at the origin is replaced by a smooth bouncing resulting from the centrifugal potential $K_{\psi}q^{-2}$. 

\subsection{Semi-classical portraits}
\label{semclassport}
By semi-classical portraits of quantum states and observables we mean representations of these objects as functions on the classical  phase space  \cite{zachos05}. In the present context, the quantum states and their dynamics have phase space representations through their \textit{ACS} or \textit{lower} symbols. Thus the ACS symbol of  $|\phi\rangle$ is defined as 
\begin{equation}
\label{Phisym} 
\Phi(q,p)= \frac{\langle q,p|\phi\rangle}{\sqrt{2\pi}}\,,
\end{equation}
with the associated probability distribution on phase space, resulting from the resolution of the identity and given by
\begin{equation}
\label{rhophi} 
\rho_{\phi}(q,p)=\dfrac{1}{ c_{-1}}|\Phi(q,p)|^{2} \,.
\end{equation}
Having at our disposal the (energy) eigenstates of some quantum Hamiltonian $\hat \sfH$, for instance the affine quantized $A_\sfH$ of a classical Hamiltonian $\sfH(q,p)$, it is particularly instructive to compute (and draw) the time evolution of the distribution \eqref{rhophi} defined as
\begin{equation}
\label{rhophiev} 
\rho_{\phi}(q,p,t):=\dfrac{1}{2\pi c_{-1}}|\langle q,p|e^{-\ii \hat \sfH t}|\phi\rangle|^{2} \,.
\end{equation}

As explained in \ref{covintquant}, the quantization map $f\mapsto A_f$ is completed with a semi-classical portrait encapsulated by the  lower symbol $\check f (q,p)$ of the operator $A_f$. This new  function  is defined as the ACS expected value of $A_f$ 
\begin{equation}
\label{expecAf}
\check f (q,p) = \lg q,p |A_f| q,p\rg \,.
\end{equation}
The explicit form of \eqref{expecAf} is given in \eqref{afflowsymb}. It  amounts to calculate the local average value of the original  $f(q,p)$  with respect to the probability distribution \eqref{rhophi} with $|\phi\rg= |q^{\prime},p^{\prime}\rg$.

%From \eqref{quantfaff} and supposing that inverting the order of the integrals is legitimate here, we obtain
%\begin{align}
%\nn
%\check{f}(q,p)= \frac{1}{\sqrt{2\pi}c_{-1}} \int_{0}^{\infty} \frac{\ud q^{\prime}}{qq^{\prime}} \int_{0}^{\infty} \int_{0}^{\infty} \ud x\ud x^{\prime}  \left[ e^{-\ii p(x-x^{\prime})}
%\,F_p(q^{\prime},x^{\prime}-x) \cdot \right.
%\\
%\label{afflowsymb}
%\left. \psi\left(\frac{x}{q}\right)\,
%\psi\left(\frac{x}{q^{\prime}}\right)\,\psi\left(\frac{x^{\prime}}{q}\right)\,
%\psi\left(\frac{x^{\prime}}{q^{\prime}}\right) \right]\,,
%\end{align}
%where $F_p$ stands for the partial inverse Fourier transform
%\begin{equation}
%\label{parcoure} F_p(q,x)=
%\frac{1}{\sqrt{2\pi}}\int_{-\infty}^{+\infty} \ud p e^{-\ii px} f(q,p)\, .
%\end{equation}

\begin{figure}[htb!]
\begin{center}
\includegraphics[width=2in]{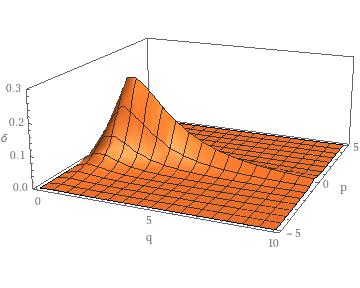}
\includegraphics[width=2in]{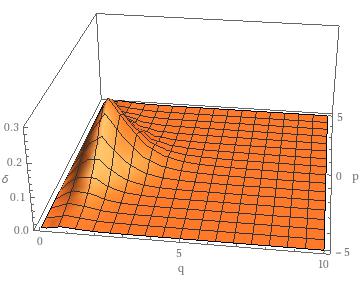}
\caption{ 3d representation, for different values of $\nu$, of  the regularized Dirac $\delta$ at the origin with the  choice of the rapidly decreasing  fiducial function \eqref{fctpsi1}. The figure on the left is for $\nu=2$ and the figure on the right is for $\nu=4$.}
\label{napdireg1}
\end{center}
\end{figure}

\begin{figure}[htb!]
\begin{center}
\includegraphics[width=4in]{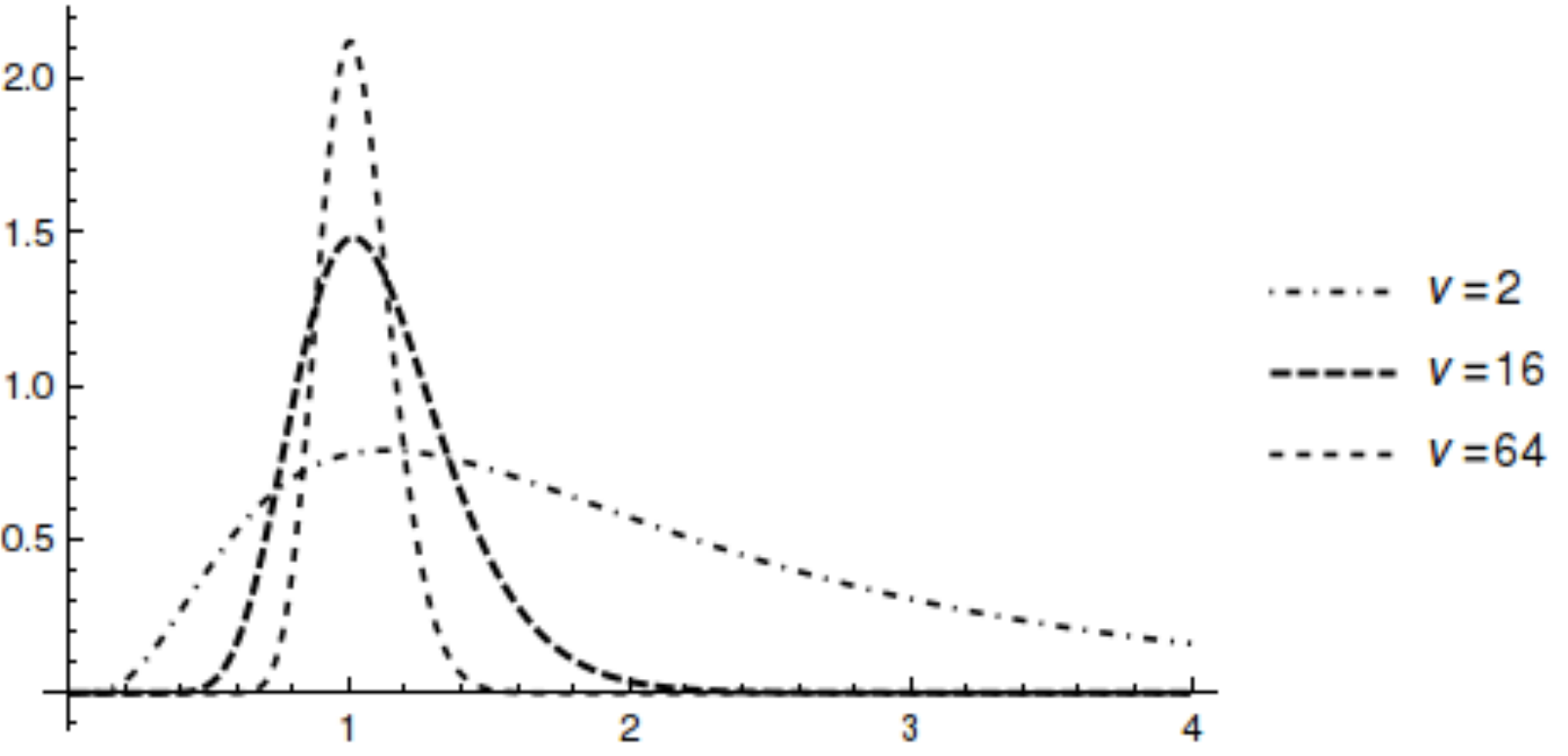}
    \caption{ Fiducial function \eqref{fctpsi1} for different $\nu$. As $\nu$ grows, it approximates to the Dirac delta.}
\label{napdireg}
\end{center}
\end{figure}

As a first example, let us calculate with real $\psi$ the lower symbol of the Dirac delta localised at $(q_0,p_0)$. We find
\begin{equation}
\label{lowsymdel1}
\check \delta_{(q_0,p_0)} = \frac{\vert\lg q,p|q_0,p_0\rg\vert^2}{2\pi c_{-1}}
= \frac{1}{2\pi c_{-1} q q_0}\left\vert \int_0^{\infty}\ud x \, e^{-\ii (p-p_0)x}\, \psi\left(\frac{x}{q}\right)\, \psi\left(\frac{x}{q_0}\right)\right\vert^2\, . 
\end{equation}
Thus we get  a new probability distribution on the phase space, centred at $(q_0,p_0)$, which regularises the original Dirac probability distribution. In Figure~(\ref{napdireg1})  is shown the shape of this regularized delta at the origin, with the following choice of rapidly decreasing fiducial function
\begin{equation}
\label{fctpsi1}
\psi_{\nu}(x) = \left( \frac{\nu}{\pi} \right)^{1/4} \frac{1}{\sqrt{x}} \exp \left[-\frac{\nu}{2} \left(\ln x - \frac{3}{4 \nu} \right)^2  \right]\,.
\end{equation}
The above real function, which is nothing but the square root of a Gaussian distribution on the real line with variable $y=\ln x$, centered at $y=3/4\nu$ ($x=e^{\frac{3}{4\nu}}$), and with variance $1/\nu$, verifies $c_{-2}(\psi_\nu)=1$, $c_0(\psi_\nu)=c_{-1}(\psi_\nu)$, and more generally
$$
c_\gamma(\psi_\nu)= \exp \left[\frac{(\gamma+2)(\gamma-1)}{4\nu} \right]\,.
$$
As $\nu\to\infty$, the  function \eqref{fctpsi1} approaches a Dirac peak. More precisely, it is shown in Fig.~(\ref{napdireg}) that as $\nu$ grows, this function smoothly concentrates around $\delta(x-1)$, which is the  position eigendistribution for  $x=1$. Conversely, as  $\nu$ goes to $0$, \eqref{fctpsi1} tends to $0$, which illustrates the total lack of information about the $x$ position. Through these features, one can understand one of the aspects of ACS quantization, which is  to smear the classical system variables.

%For functions $f$ depending on $q$ only, expression \eqref{afflowsymb} simplifies to a lower symbol  depending on $q$ only:
%\begin{align}
%\nn
%\check{f}(q)&= \frac{1}{c_{-1}}\int_0^{\infty}\frac{\ud
%q^{\prime}}{qq^{\prime}}\, f(q^{\prime}) \int_0^{\infty}\ud
%x\,\left(\psi\left(\frac{x}{q}\right)\right)^2\,\left(\psi\left(\frac{x}{q^{\prime}}\right)\right)^2
%\\
%\label{lowfq1} &= \frac{1}{c_{-1}} \left\lg \frac{1}{q}\psi^2\left(\frac{\cdot}{q}\right)\Big|f\ast_{\mathrm{aff}}\psi^2\right\rg\, ,
%\end{align}
%where the convolution $\ast_{\mathrm{aff}}$ is defined in \eqref{convdef}. 
Other examples, which are particularly relevant to the content of the present paper and whose calculations are developed in  \ref{proofresunit}, are given below.

\subsubsection*{Lower symbol of powers of $q$}  It is given with the same power up to a constant factor
\begin{equation}
\label{powq} 
q^{\beta} \mapsto  \check{q^{\beta}}=
\frac{c_{\beta-1}c_{-\beta-2}}{c_{-1}} \, q^{\beta}\, . 
\end{equation}

\subsubsection*{Lower symbols of momentum, kinetic energy, and product $qp$}

Calculated with real $\psi$, they read respectively 
\begin{equation}
\label{symbp}
p \mapsto \check{p}= p\, ;
\end{equation}
\begin{equation}
\label{symbp2} 
p^2 \mapsto \check{p^2}= p^2 + \frac{\mathrm{c(\psi)}}{q^2}\, , \quad \mathrm{c(\psi)} = \int_0^{\infty}\left(\psi^{\prime}(x)\right)^2\,\left(1+\frac{c_0}{c_{-1}}x\right)\,\ud x\,;
\end{equation}
\begin{equation}
\label{symbp2A} 
qp \mapsto \check{qp}= \frac{c_0 c_{-3}}{c_{-1}}qp\, .
\end{equation}

\section{A short reminder of Lagrangian and Hamiltonian formalism} 
\label{lagham}

The motion on the half-line of a particle of mass $m$ is described by the Lagrangian:
\begin{equation}
\label{lag1d}
\sfL(q,\dot q, t) = \frac{m\dot q^2}{2} - V(q)\, , 
\end{equation}
where $q>0$ and $V$ is the potential. The corresponding  Lagrange equation  reads
\begin{equation}
\label{lageq}
0= \frac{d}{dt}\frac{\partial \sfL}{\partial \dot q} - \frac{\partial \sfL}{\partial q}= m\ddot q + V^{\prime}(q)\, .
\end{equation}
One passes  to the Hamiltonian formalism through 
the momentum
\begin{equation}
\label{momentum}
p := \frac{\partial \sfL}{\partial \dot q} = m\dot q\, , 
\end{equation}
and the corresponding Hamiltonian
\begin{equation}
\label{hamilton}
\sfH= p\dot q -\sfL = \frac{p^2}{2m} + V(q)\, . 
\end{equation}
In the sequel, we will be interested in potentials of the form of superposition of powers of $q$
\begin{equation}
\label{potential}
V(q) = \int_{-\infty}^{+\infty} w(\beta)\, q^{\beta}\,\ud \beta\,, 
\end{equation}
where the weight function, usually with bounded support, can be extended to a distribution.

Due to the time independence of the Hamiltonian, energy =$\,\sfH$ is conserved and phase space trajectories on the half-plane are determined by initial conditions $(q_0,p_0)$ as 
\begin{equation}
\label{phsptrj}
\frac{p^2}{2m} + V(q) = E:= \frac{p_0^2}{2m} + V(q_0)\, .
\end{equation}
The flow along them is described by Hamilton equations
\begin{equation}
\label{hameq}
\dot q = \lbrace q,\sfH \rbrace= \frac{p}{m} \quad, \quad \dot p  = - \lbrace p,\sfH \rbrace= -V^{\prime}(q)\, .
\end{equation}

\section{ACS quantization of dynamics on half-line}
\label{acshalfline}

Having in hand the formulas established in the previous section, it is is straightforward to establish the quantum version of the Hamiltonian \eqref{hamilton}-\eqref{potential}:
\begin{equation}
\label{qhamilton}
A_{\sfH}= \frac{P^{2}}{2m}+\frac{K_{\psi}}{2Q^{2}} + A_V \, ,
\end{equation}
with
\begin{equation}
\label{qpot}
A_V= \int_{-\infty}^{+\infty} w(\beta)\, \frac{c_{\beta-1}}{c_{-1}}\, Q^{\beta}\,\ud \beta\, . 
\end{equation}
In the sequel, we suppose that the weight function $w(\beta)$ and the fiducial vector are chosen such that the quantum Hamiltonian $A_{\sfH}$ is essentially self-adjoint: quantum dynamics does not depend on a choice of boundary conditions at the origin of the half-line. 

Concerning our choice of fiducial vector, we  consider  two other options, in addition to our previous choice \eqref{fctpsi1}. 
The most immediate is to pick one of the elements of the well-known orthonormal basis of $L^{2}(\mathbb{R}_{+}^{\ast},\mathrm{d}x)$   built from Laguerre polynomials \cite{magnus66},
\begin{equation}
\label{LagOB}
 e^{(\alpha)}_n(x) := \sqrt{\frac{n!}{(n+\alpha)!}}\, e^{-\frac{x}{2}}\, x^{\frac{\alpha}{2}}\, L_n^{(\alpha)}(x) \,, \quad \int_{0}^{\infty}\eal_n(x)\, \eal_{n^{\prime}}(x) \ud x = \delta_{n n^{\prime}}\,, 
\end{equation}
where $\alpha > -1$ is a free parameter, and $(n+\alpha)! = \Gamma(n+\alpha + 1)$. Actually,  since we wish to work with  functions which, with a certain number of their derivatives,  vanish at the origin, the parameter $\alpha$ should  be imposed to be larger than  some  $\alpha_0 >0$. On the other hand, for a general $n$, the expression of the constants $c_{\gamma}$ appears quite involved \cite{gradryz07}:
\begin{equation}
\label{cgamalg}
c_{\gamma}(\eal_n)= \frac{\Gamma(\alpha -\gamma -1)}{\Gamma(\alpha +1)}\, \frac{1}{n!}\, \frac{\ud^n}{\ud h^n} \,\frac{{}_2F_1 \left( \frac{\alpha -\gamma-1}{2}, \frac{\alpha-\gamma}{2}; \alpha +1; \frac{4h}{(1+h)^2}\right)}{(1+h)^{\alpha-\gamma-1}(1-h)^{\gamma+2}}
\Big\vert_{h=0}\, ,
\end{equation}
which is valid for $\alpha >\gamma + 1$. 

The second option is the normalised function in $\LL$ \cite{berdagama14}: 
\begin{equation}
\label{fidpsi3}
\psi(x)=\psi^{\nu,\xi}(x)= \frac{1}{\sqrt{2 x \, K_0(\nu)}}e^{-\frac{ \nu}{4} \left(\xi x +\frac{1}{\xi x}\right)} \,,
\end{equation}
with $\nu >0$ and $\xi>0$. Here and in the following, $K_r(z)$ denotes the modified Bessel functions \cite{magnus66}. Actually, we only deal with ratios of such functions throughout. Whence we adopt the convenient notation 
\begin{equation}
\label{xibessel}
\xi_{rs} = \xi_{rs}(\nu)=  \frac{K_r(\nu)}{K_s(\nu)}= \frac{1}{\xi_{sr}}\, .
\end{equation}
One attractive feature of such a notation is that $\xi_{rs}(\nu) \sim 1$ as $\nu \to \infty$ (we recall that the asymptotic behavior at large argument $\nu$ is $K_r(\nu) \sim e^{-\nu}\sqrt{\pi/(2\nu)}$, whereas at small $\nu\ll \sqrt{r+1}$, $K_r(\nu) \sim (1/2)\Gamma(r)(2/\nu)^r$ for $r>0$ and $K_0(\nu) \sim -\ln(\nu/2) -\gamma$). 
We notice that $\psi^{\nu,\xi}(x)$ falls off with all its derivatives at the origin and at the infinity. Normalization constant and other integrals involving the function $\psi^{\nu,\xi}$ are easily obtained thanks to the formula \cite{gradryz07}
\begin{equation}
\int_0^\infty x^{a-1} e^{-c x-b/x} \ud x = 2 \left( \frac{b}{c} \right)^{a/2} K_{a}(2 \sqrt{b c})\, , 
\end{equation}
$\forall \, a, b, c \in \mathbb{C}, \, \Re(b) >0, \Re(c) >0$. With such a fiducial vector the integrals $c_{\gamma}$ read as
\begin{equation}
c_\gamma\left(\psi^{\nu,\xi}\right) = \xi^{\frac{\gamma}{2}+2}\,\frac{K_{-\gamma-2}( \nu)}{K_0(\nu)}= \xi^{\frac{\gamma}{2}+2}\,\xi_{-\gamma-2, 0}\, .
\end{equation}

With these fiducial functions, we have two free parameters $\xi$ and $\nu$ (besides the scaling parameter $\kappa$). Hence some freedom is left to us to give ratios $c_{\gamma}/c_{\gamma^{\prime}}$ the value we wish, an opportunity we use in the first example (next Section).  

%Fixing the parameter $\xi$ allows to 
%\begin{equation}
%\xi = \frac{K_{-1}(\nu)}{K_{-3}(\nu)}= \frac{\xi_{-1,-3}}{\sqrt{2}}\,.
%\end{equation}
%The interest of this choice will become apparent with the semi-classical analysis of the operator $A_q$. Therefore $\nu$ remains the only free parameter. 
%With this, we find the relation
%\begin{equation}
%\left(\frac{c_{-1}}{c_{1}}\right)^{2}= 2\,.
%\end{equation}

\section{First example: quantum breathing of a massive sphere}
\label{msphere}

\begin{figure}[htb!]
	\centering
	\begin{minipage}{.5\textwidth}
		\centering
		\includegraphics[width=.8\linewidth]{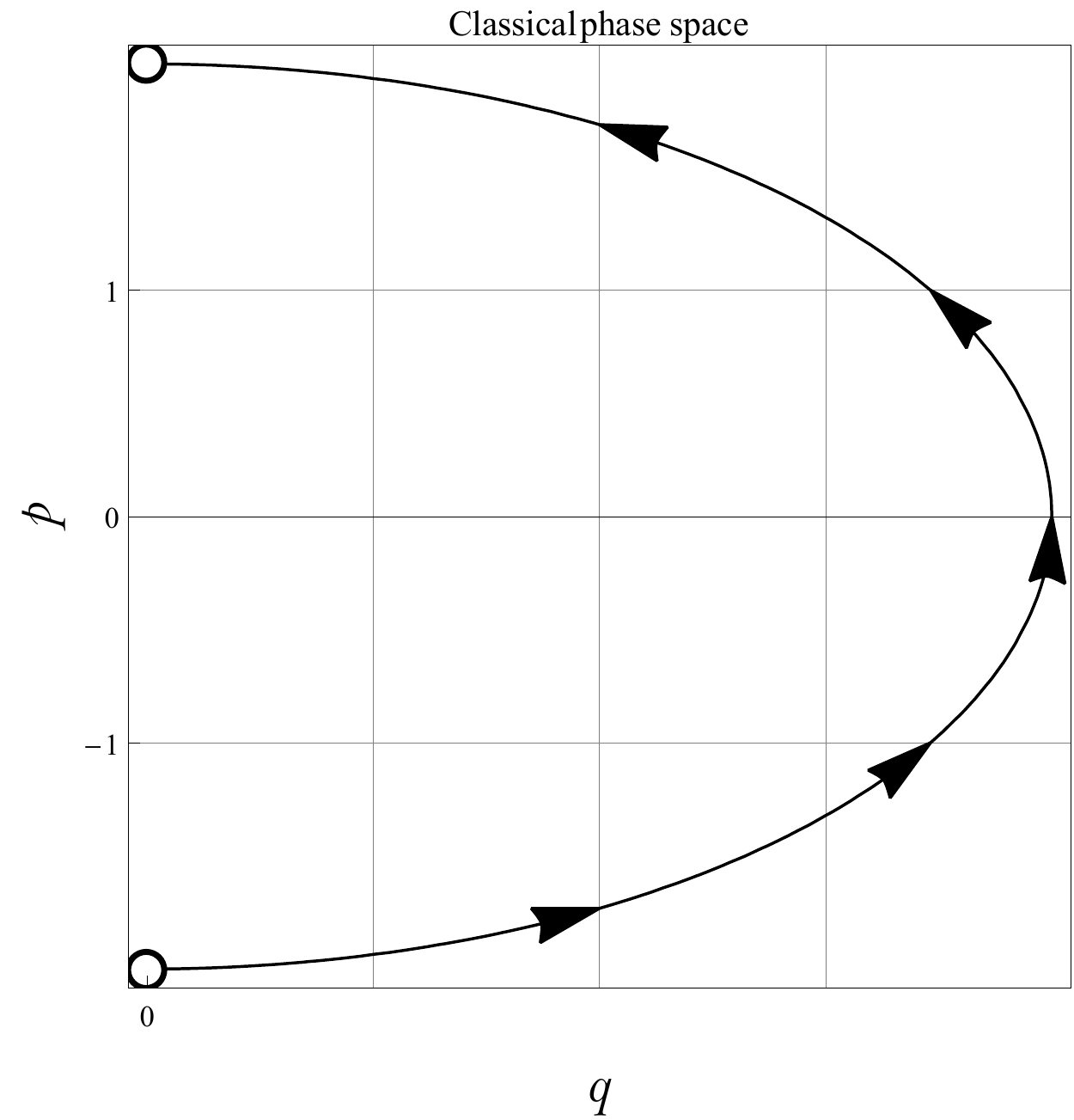}
		\subcaption{Classical trajectory}
		\label{phtrajcd1} 
	\end{minipage}%
	\begin{minipage}{.5\textwidth}
		\centering
		\includegraphics[width=.8\linewidth]{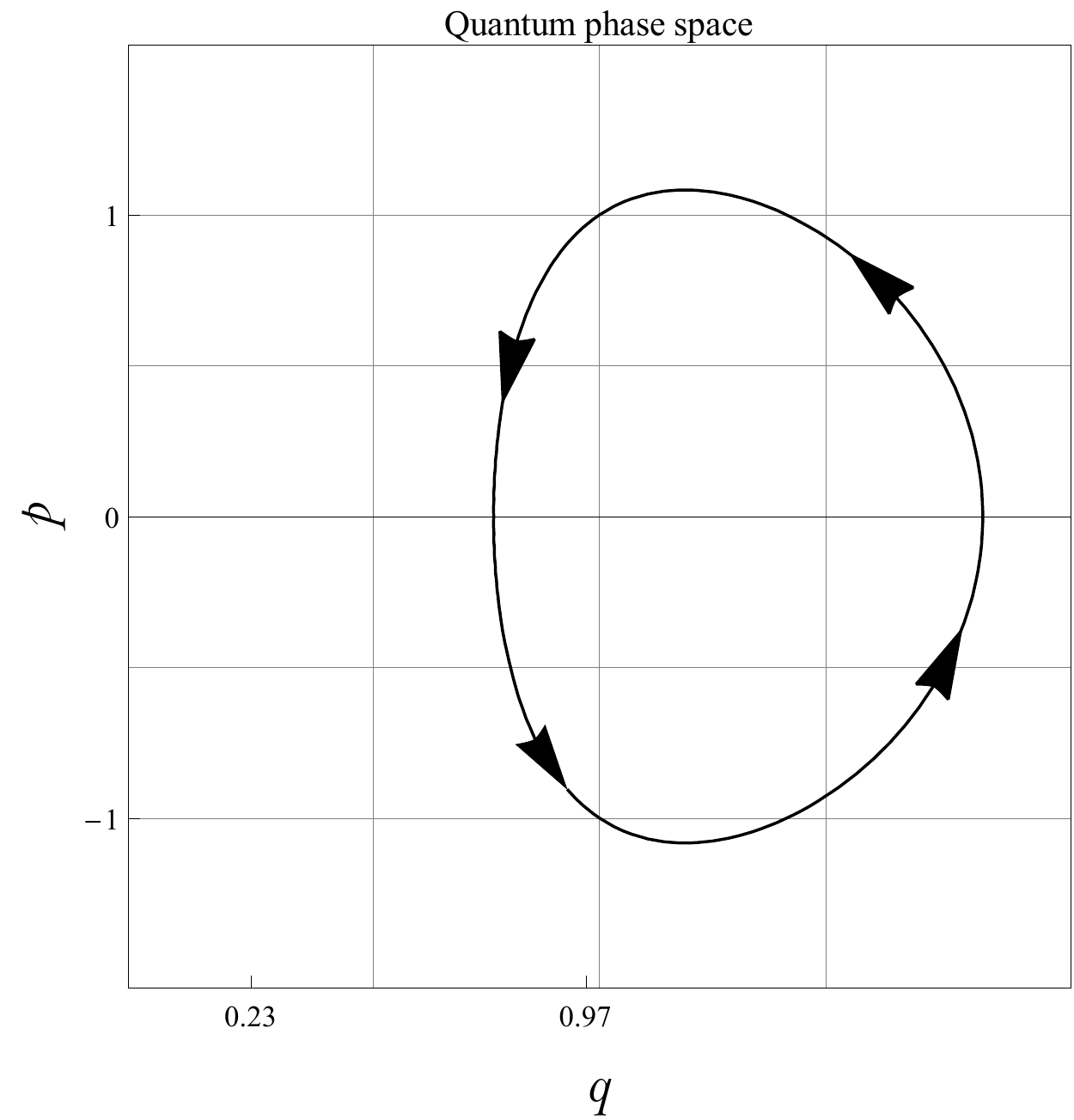}
		\subcaption{Semi-classical trajectory}
		\label{scphtrajcd1}
	\end{minipage}
	\caption{\textbf{Figure \eqref{phtrajcd1}} is an example of phase space trajectory in the positive half-plane defined by the equation $E= p^2/2 +k q^2/2$ with $E= 2$ and $k=1$. The reflection at the origin produces the momentum discontinuity $-p_0 \mapsto p_0$.  \textbf{Figure \eqref{scphtrajcd1}} is an example of ACS semiclassical regularised phase space trajectory in the positive half-plane defined by the equation \eqref{schamilpotsphA}  with $E= 2$, $\tilde k=1$, and  $\tilde K= 1$. The latter choices for  $\tilde K$ and $\tilde k$ are easily made possible thanks to a suitable fixing of parameters of the fiducial vector, as was stressed at the end of the previous section.  The classical reflection has become a smooth bouncing near the origin. }
\end{figure}

Let us consider an isotropic  medium with constant mass density $\rho_0$.  This implies that the ball of center $O$  and  radius $q$ has a total mass equal to
\begin{equation}
\label{cstdensA}
M(q)= \frac{4\pi}{3}\, \rho_0\, q^3 \, .
\end{equation}
Gauss theorem allows to determine easily  the gravitational vector field acting on a test mass at the surface of the ball. Hence, 
the Newton equation for a test particle, mass $m$, at the surface of the sphere of radius $q$ reads as
\begin{equation}
\label{newtestA}
m \ddot{q} = -\frac{G m M(q)}{q^2}=  -m\,\frac{4\pi G}{3}\, \rho_0\, q\equiv -k\,  q \, , 
\end{equation}
where $G
%= 6.67384 \times 10^{-11}\, \mathrm{m}^3\, \mathrm{kg}^{-1} \,\mathrm{s}^{-2}
$ is the universal gravitational constant. The Hamiltonian is the same as the one for the half-harmonic oscillator  \cite{griffiths05}, that is, whose the motion is restricted to  the half-line. A physical interpretation of this could be a spring that can be stretched from its equilibrium position but not compressed. With the choice $m= 1$\,kg,  
\begin{equation}
\label{hamilpotsphA}
\sfH = \frac{p^2}{2} + k\frac{q^2}{2}\, , \quad p=\dot q\, \quad q>0 \, . 
\end{equation}
An example of phase space trajectory, a truncated circle, is given in Figure \ref{phtrajcd1}. 

%\begin{figure}[!htb]
%\begin{center}
%\includegraphics[scale=0.5]{figuras_novas/classical_phase_spaceI}\caption{An example of phase space trajectory in the positive half-plane defined by the equation $E= p^2/2 +k q^2/2$ with $E= 1=k$ with $q\equiv x$ and $p\equiv y$.}
%\label{phtrajcd1} 
%\end{center}
%\end{figure}

According to \eqref{qhamilton}, the ACS quantization of this classical dynamics yields the quantum Hamiltonian
\begin{equation}
\label{qhamiltonS1A}
A_{\sfH}= \frac{P^{2}}{2}+\frac{\hbar^2}{2}\frac{K_{\psi}}{Q^{2}} +  \frac{k}{2}\,\frac{c_1}{c_{-1}}Q^2\, , 
\end{equation}
in which the presence of the Planck constant  is due to the fact that we have to take into account  the physical dimensions of the phase space variables $(q,p)$  and consistently replace in \eqref{quantfaff} the measure $\ud q \,\ud p$ by $\ud q \,\ud p/\hbar$. Passing to the lower symbol of the equation \eqref{qhamiltonS1A} through formulas given in \eqref{powq} and \eqref{symbp2} at constant energy $A_{\sfH}= E$ yields the semi-classical correction to \eqref{hamilpotsphA}
\begin{equation}
\label{schamilpotsphA}
E= \frac{p^2}{2} +\frac{ \hbar^2}{2}\,\frac{c(\psi)}{q^2} + \frac{k}{2} \frac{c_1\,c_{-4}}{c_{-1}}q^2\equiv  \frac{p^2}{2} + \frac{\tilde K}{q^2} +
 \frac{ \tilde k}{2} q^2\, ,
\end{equation}
where $c(\psi)$ is defined in \eqref{symbp2}.

The presence of the centrifugal potential in equation \eqref{schamilpotsphA}, of purely quantum origin, allows to eliminate the singularity due to the reflection by creating a smooth bouncing as it is illustrated by Figures~(\ref{scphtrajcd1}).

Note that there is a modification of the oscillator strength $k$ which becomes $\tilde k$. If one considers this fact as a problem, the ``renormalised" $\tilde k$ can be  made arbitrarily close  to $k$ by choosing in a suitable way the parameters present in the expression of the fiducial $\psi$. For instance, with the choice of fiducial \eqref{fidpsi3}, we have $\tilde{k} = \xi^{4} \, \xi_{30} \,\xi_{2-1}$ and with $\xi=1$,   the product $\xi_{30} \, \xi_{2-1}$ becomes rapidly closer to $1$, as shown in the Figure~(\ref{xi_-4,0}). On the other hand, one could decide that what is measured is not $k$, which pertains to the classical model, viewed as incomplete because ``classical'', but rather the ``effective" $\tilde k$, viewed as more  ``realistic'' since we suppose that the quantum model supersedes the classical one. This might open a debate analogous to that one arising from the distinction between bare mass and dressed or effective mass in Quantum Field Theory. 

\begin{figure}[!htb]
	\begin{center}
		\includegraphics[scale=0.5]{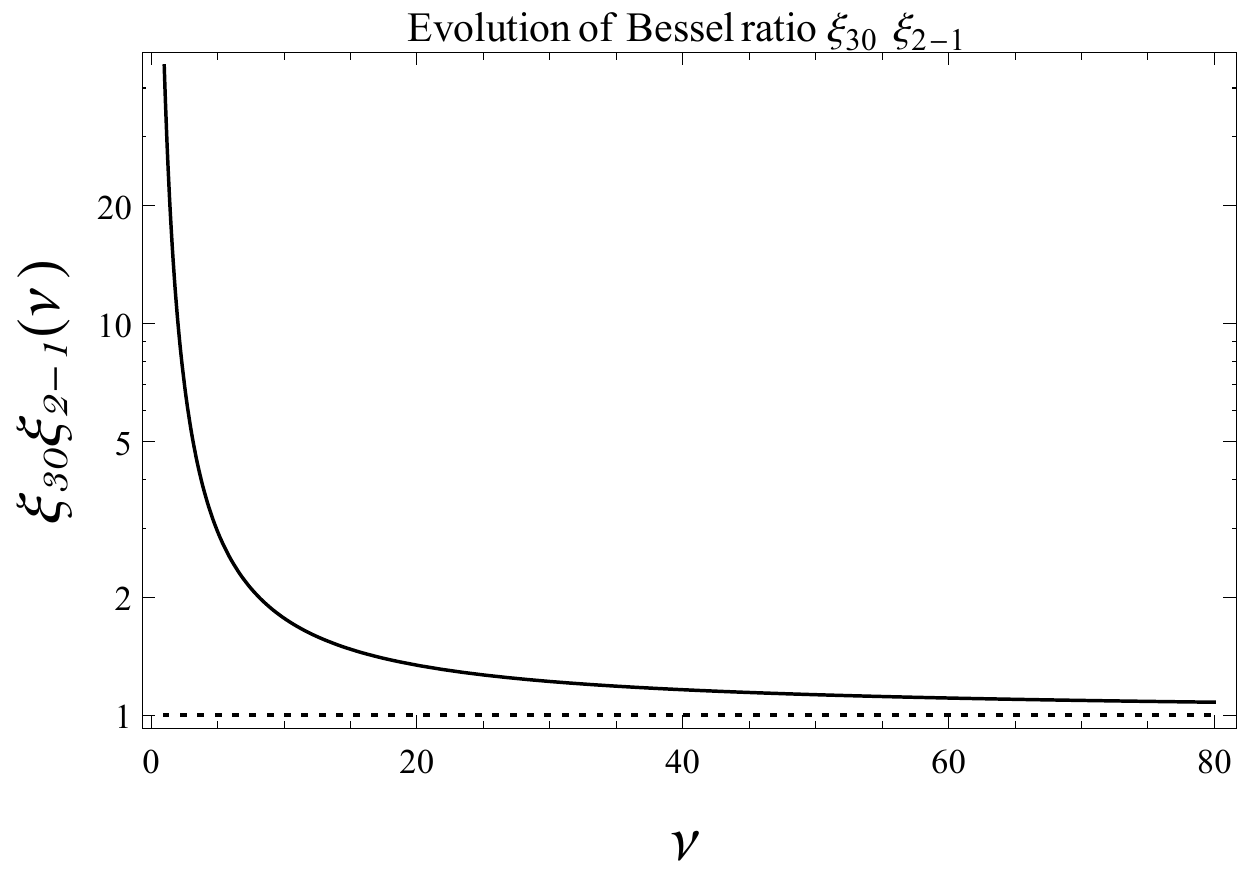}
		\caption{$\xi_{30} \, \xi_{2-1}$ rapidly becomes closer to $1$ for larger $\nu$. }
		\label{xi_-4,0}
	\end{center}
\end{figure}

%\begin{figure}[!htb]
%\begin{center}
%\includegraphics[scale=0.5]{figuras_novas/quantum_phase_space}
%\caption{An example of semiclassical phase space trajectory in the positive half-plane defined by the equation \eqref{schamilpotsphA}  with $E= 1=\tilde k$,  $\tilde K= 0.05$, with $q\equiv x$ and $p\equiv y$.}
%\label{scphtrajcd1}
%\end{center}
%\end{figure}

The eigenvalues $E_n$ and eigenfunctions $\phi_n$ of equation \eqref{qhamiltonS1A} in its operator form can be found by solving the eigenvalue equation
\begin{equation} 
\label{eigvaeq1}
\frac{1}{2}\left(-\hbar^2\partial^2_x + \frac{\hbar^2 K_{\psi}}{x^2} + \tilde k\, x^2 \right) \phi_n = E_n \phi_n \, .
\end{equation}
Defining the quantities
\begin{equation}
\mu = \frac{1}{2}\sqrt{1+4K_{\psi}} \, , \quad \lambda = \frac{1}{2\hbar^2}\left( \frac{k^2}{2}\right)^{\frac{1}{4}} \, ,
\end{equation}
the solutions are a combination of exponentials and associated Laguerre polynomials as in the following
\begin{equation} \label{eigfun1}
\phi_n (x) = 2^{\frac{1}{2} (\mu +1)} x^{(\mu + \frac{1}{2})} e^{-\lambda x^2} L_{n}^{\mu}\left( 2\lambda x^2 \right) \, ,
\end{equation}
with $n \in \N$, and the eigenvalues are given by
\begin{equation} 
\label{eigva1}
E_n = 2\hbar^3 \lambda \left(2n + \mu + 2 \right) \,.
\end{equation}
A similar model was  analysed in full analytical and numerical details in the article \cite{berdagama14} devoted to the study of gravitational singularities in the case of  the Robertson- Walker metric coupled to a perfect fluid.

\section{Second example: quantum bouncing of  charged sphere}
\label{chsphere}

\begin{figure}[htb!]
	\centering
	\begin{minipage}{.5\textwidth}
		\centering
		\includegraphics[width=.8\linewidth]{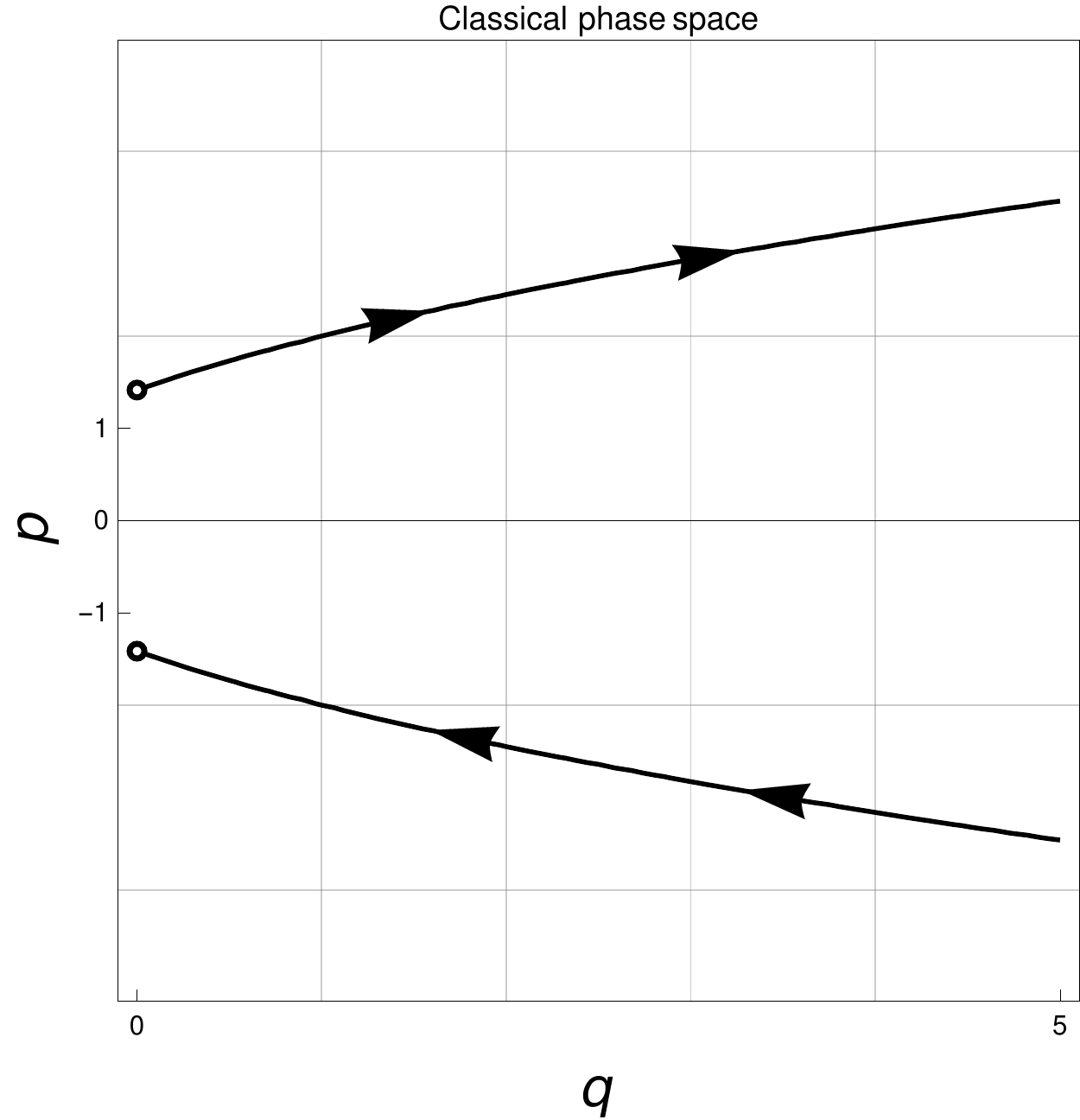}
		\subcaption{Classical trajectory}
		\label{phtraj1} 
	\end{minipage}%
	\begin{minipage}{.5\textwidth}
		\centering
		\includegraphics[width=.8\linewidth]{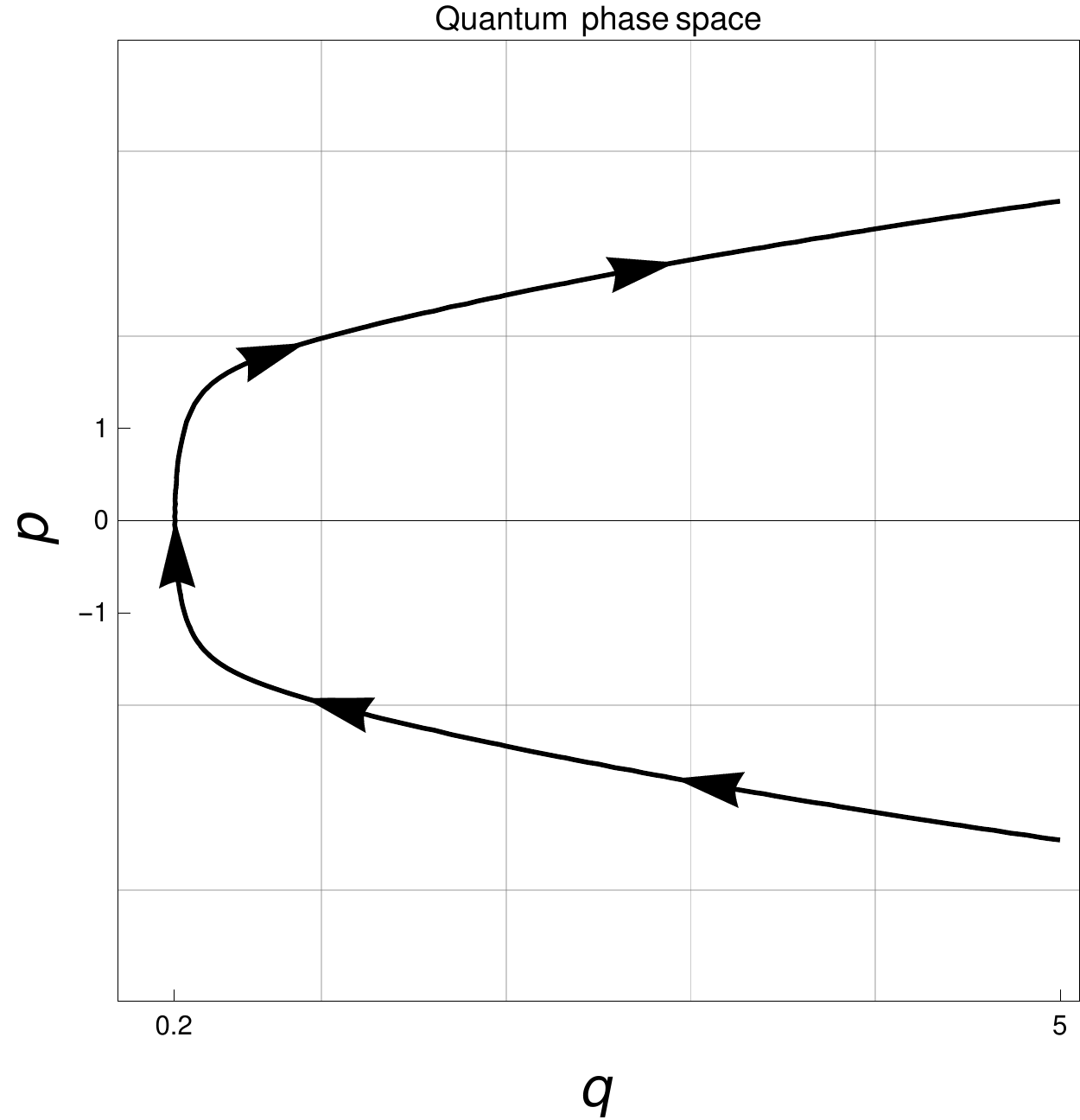}
		\subcaption{Semi-classical trajectory}
		\label{scphtraj1}
	\end{minipage}
	\caption{Figure \eqref{phtraj1} is an example of phase space trajectory in the positive half-plane defined by the equation $E= p^2/2 -k q$ with $E= 1=k$. Figure \eqref{scphtraj1} is an example of semiclassical phase space trajectory in the positive half-plane defined by the equation \eqref{schamilpotsph}  with $E= 1=\tilde k= \tilde K$.}
\end{figure}

Let us consider an isotropic negatively charged  insulating medium whose the density of charge varies as $1/q$ at distance $q$ (don't confuse with a charge!) of the  symmetry center $O$.  This means that the ball of center $O$  and  radius $q$ has a total charge equal to
\begin{equation}
\label{cstdens}
\mathfrak{Q}= -k_s\, q^2\,, \quad k_s > 0\, .
\end{equation}
The Newton equation for a test positive unit charge  and unit mass, at distance $q$ of the center,  reads as
\begin{equation}
\label{newtest}
 \ddot{q} = \frac{\mathfrak{Q}}{4\pi \epsilon_0 q^2 }= -\frac{k_s}{4\pi \epsilon_0}\equiv -k\, , 
\end{equation}
where the radial force or electric field acting on the test charge is determined by using the Gauss theorem. 
The corresponding  Hamiltonian is given by
\begin{equation}
\label{hamilpotsph}
\sfH = \frac{p^2}{2} - kq\, ,
\end{equation}
which corresponds to the weight function $w(\beta) \propto - \delta(\beta -1)$ in \eqref{potential}. An example of phase space trajectory, a truncated parabola, is given in Figure~(\ref{phtraj1}). 

%\begin{figure}[!htb]
%\begin{center}
%\includegraphics[scale=0.5]{figuras_novas/classical_phase_spaceII}
%\caption{An example of phase space trajectory in the positive half-plane defined by the equation $E= p^2/2 -k q$ with $E= 1=k$ with $q\equiv x$ and $p\equiv y$.}
%\label{phtraj1}
%\end{center}
%\end{figure}

According to \eqref{qhamilton}, the ACS quantization of this classical dynamics yields the quantum Hamiltonian
\begin{equation}
\label{qhamiltonS1}
A_{\sfH}= \frac{P^{2}}{2}+\frac{\hbar^2}{2}\frac{K_{\psi}}{Q^{2}} - k \frac{c_0}{c_{-1}}Q\, , 
\end{equation}
in which the insertion  of $\hbar$ has same justification as  for \eqref{schamilpotsphA}. Passing to the lower symbol of the equation \eqref{qhamiltonS1}, through formulas given in \eqref{powq} and \eqref{symbp2} at constant energy $A_{\sfH}= E$ yields the semi-classical correction to \eqref{hamilpotsph}
\begin{equation}
\label{schamilpotsph}
E= \frac{p^2}{2} + \frac{\hbar^2}{2}\,\frac{c(\psi)}{q^2} - k \frac{c_0\,\,c_{-3}}{c_{-1}}q\equiv  \frac{p^2}{2} + \frac{\tilde K}{q^2} - \tilde k q\, .
\end{equation}

Again, the presence of the centrifugal potential, of purely quantum origin, allows to eliminate the singularity by creating a smooth bouncing as it is illustrated by Figure~(\ref{scphtraj1}). Nevertheless, for this case, we cannot obtain analytical solutions for the eigenvalue equation derived from \eqref{qhamiltonS1} in its operator form
\begin{equation}
\frac{1}{2} \left( -\hbar^{2} \partial^{2}_{x} + \hbar^2\frac{K_{\psi}}{x^{2}} - k \frac{c_0}{c_{-1}}x \right) \phi_{n} = E_{n} \phi_{n} \, , 
\end{equation} 
and, therefore, the time evolution of a state could be only calculated numerically. 
%\begin{figure}[!htb]
%\begin{center}
%\includegraphics[scale=0.5]{figuras_novas/quantum_phase_spaceII}
%\caption{An example of semiclassical phase space trajectory in the positive half-plane defined by the equation \eqref{schamilpotsph}  with $E= 1=\tilde k= \tilde K$ with $q\equiv x$ and $p\equiv y$.}
%\label{scphtraj1}
%\end{center}
%\end{figure}
\section{Dust in  cosmology}
\label{dust}
In our third example, the most expanded one in this paper, we deal with the simple model of dynamics of dust in Newtonian cosmology that is presented  by Mukhanov in Chapter 1 of his book \cite{mukhanov05}.                   
We consider a sphere of radius $q(t)$ in an infinite, expanding, homogeneous and isotropic universe filled with \textit{dust}, i.e. a matter with negligible pressure compared with its energy density. Newtonian gravity is applicable in the case of weak gravity and not too large radius. Also using Gauss theorem, one ignores  the gravitational effect on a particle within the sphere due to the matter outside the sphere, a feature which can be also justified within the framework of general relativity (Jebsen-Birkhoff theorem \cite{synge60}). Therefore, the  Newton equation applied to a probe mass $m$ located at the surface of the sphere reads
\begin{equation}
\label{dustnewton}
m\ddot q= - \frac{GmM}{q^2}\, , 
\end{equation}where $M$ is the time-independent mass of the sphere. Deleting the probe mass,  the corresponding Hamiltonian is Kepler-like,
\begin{equation}
\label{kepler}
\sfH= \frac{p^2}{2} - \frac{k}{q}\, , \quad k= GM\,. 
\end{equation}
According to \eqref{qhamilton}, the ACS quantization of this classical model gives the quantum Hamiltonian:
\begin{equation}
\label{qkepler}
A_{\sfH}= \frac{P^{2}}{2}+\frac{\hbar^2}{2}\frac{K_{\psi}}{Q^{2}} -  \frac{1}{c_{-1}}\frac{k}{Q}\, . 
\end{equation}
Applying our general formulas\eqref{czechqbeta} and \eqref{pp2}, we obtain its lower symbol at constant energy $A_{\sfH}= E$ 
\begin{equation}
\label{sckepler}
E= \frac{p^2}{2} +\frac{\hbar^2}{2}\,\frac{c(\psi)}{q^2}  - \frac{k}{q}\equiv  \frac{p^2}{2} + \frac{\tilde K}{q^2} - \frac{ k}{q}
\, .
\end{equation}
with $c(\psi)$ defined in \eqref{symbp2}. It is  the semi-classical correction to \eqref{kepler}.  Note that in this case, there is no ``renormalisation" of the classical gravitational coupling $k$. 

 The  spectrum of the operator $A_{\sfH}$ is analogous to the  spectrum  of the Hydrogen atom obtained from the resolution of the radial Schr\"{o}dinger equation with non-zero angular momentum (of course, there is no degeneracy in the present model).  Hence, we have to distinguish between pure point spectrum corresponding to the bound states and the continuous spectrum corresponding to the scattering states. In the present example, bound states describe a pulsing or breathing dust sphere while scattering states correspond to a bouncing without recollapse. 

The eigenvalues $E_n$ and eigenfunctions $\phi_n$ of equation \eqref{qkepler} in the case of the bound states are given by
\begin{equation} 
	\label{eigvaeqkepler}
	\frac{1}{2}\left(-\hbar^2\partial^2_x + \frac{\hbar^2 K_{\psi}}{x^2} -  \frac{2}{c_{-1}}\frac{GM}{x}\  \right) \phi_n = E_n \phi_n \, .
\end{equation}
Redefining the parameters as
\begin{equation}
	\kappa_n^{2} =  -\frac{2E_n}{\hbar^{2}} \quad; \quad \alpha = \frac{1}{2} + \frac{1}{2} \sqrt{1+
		4K_{\psi}} \,,
	\end{equation}
	the square-integrable solutions to this equation are:
	\begin{equation} \label{eigfunkepler}
	 \phi_n (x) = N(n,\alpha) \, e^{-\kappa_n x} \left( \kappa_n x \right)^{\alpha} L^{(2\alpha -1)}_{n} \left( 2\kappa_n x \right) \, ,
	\end{equation}
	with $n \in \N$, and $N(n,\alpha)$ is the normalization factor given by the expression:
	\begin{equation}
		N(n,\alpha) = 2^{\alpha}\sqrt{\frac{\kappa_n}{(n+\alpha) \, n! \, \Gamma(2\alpha +n)}}\,.
	\end{equation} 
	 The eigenvalues of Equation \eqref{eigvaeqkepler} are given by
	\begin{equation} 
		\label{eigvakepler}
		\kappa_n =  \frac{G M}{ \hbar^{2} c_{-1}\, (n+\alpha)}\quad \Rightarrow \quad E_{n} = - \frac{G^{2} M^{2} }{2 \hbar^{2} (c_{-1})^{2} (n+\alpha)^{2}}  \,.
	\end{equation}
With this we can find the time evolution distribution function \eqref{rhophiev} by choosing  the normalized fiducial vector $|\psi\rg$ as
	\begin{equation}
	\psi(x) = \frac{9}{\sqrt{6}} \, x^{\frac{3}{2}} \, e^{-\frac{3x}{2}} \,,
	\end{equation}
	such as the constants $K_{\psi}$ and $c_{-1}$ are respectively $3/4$ (so $\alpha = 3/2$) and $1$. \\
	With this choice, the semiclassical expression of the energy \eqref{sckepler} reads
	\begin{equation}
	\label{semiclas1}
	E = \frac{p^2}{2} + \frac{9}{8} \frac{1}{q^2} - \frac{ GM}{q} \,.
	\end{equation}
%	Remark: this semiclassical expression is not corrected from the scaling induced by lower symbols, namely the fact that $\langle q, p | Q | q, p \rangle = (4/3) q \ne q$.\\

\begin{figure}[htb!]
	\centering
	\includegraphics[scale=0.3]{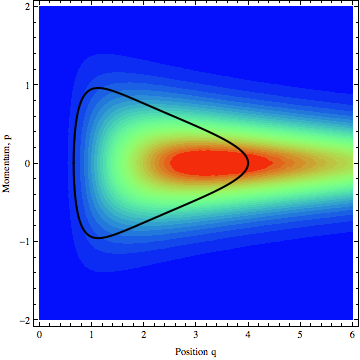}
	\includegraphics[scale=0.3]{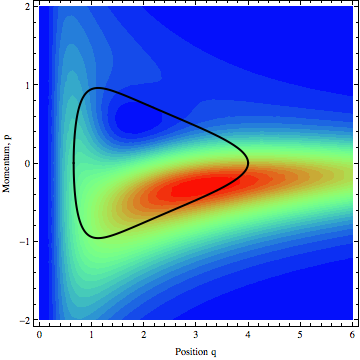}
	\includegraphics[scale=0.3]{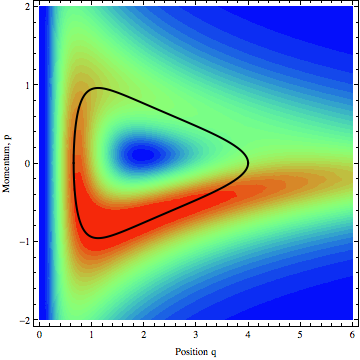}
	\includegraphics[scale=0.3]{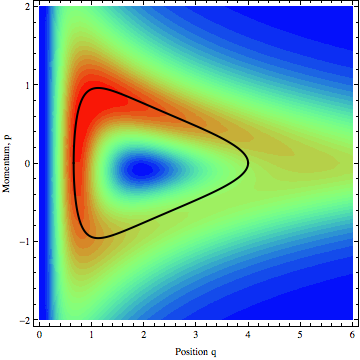}
	\includegraphics[scale=0.3]{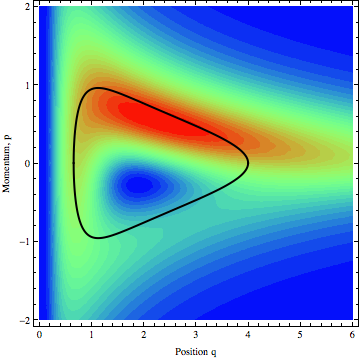}
	\includegraphics[scale=0.3]{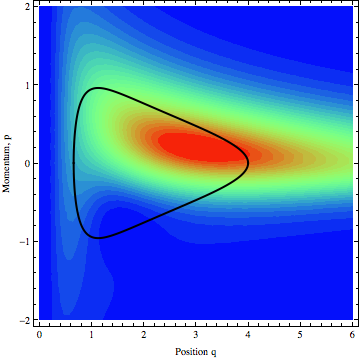}
	\caption{\label{wpakfig} Phase space representation of the quantum dynamical behavior of an initial coherent state $|q_0=4, p_0=0 \rangle$. We choose $G=\hbar=1$ and $M=2$. On each figure the thick curve represents the semi-classical trajectory due to \eqref{semiclas1} for these particular values, while increasing values of the density $\rho(q,p)$ are encoded by colors from blue to red. The different figures show the evolution of the density $\rho(q,p)$. Time is increasing from the top left to the bottom right.}
\end{figure}

Let us choose as an initial state a  coherent state $|\phi\rg = |q_0,p_0\rg$. The time-dependent probability distribution on the phase space reads:
\begin{equation}
\label{distq0p0}
\rho_{\phi}(q,p,t) = \rho_{q_0,p_0}(q,p,t)=  \frac{1}{2\pi}\vert \lg q,p| e^{-\ii A_{\sfH} t} |q_0,p_0\rg\vert^2\, .
\end{equation}	
In order to get a qualitative idea of this distribution, we project   the initial state onto the finite-dimensional subspace $\mathcal{H}_{n_{\mathrm{max}}}$ spanned by the orthonormal set of bound states $\{|\phi_n \rg\}_{0\leq n \leq n_{\mathrm{max}}}$. 
\begin{equation}
\label{projnm}
|q_0,p_0\rg \mapsto |q_0,p_0\rg_{n_{\mathrm{max}}}:= \sum_{n=0}^{n_{\mathrm{max}}} c_n(q_0,p_0) |\phi_n \rg \,, 
\end{equation}
where the coefficients $ c_n(q_0,p_0) := \lg \phi_n | q_0,p_0\rg$ are given (for a general $\alpha$) by
\begin{align}
\label{cnqp}
%\begin{split}
\nonumber c_n(q_{0},p_{0}) = \frac{9}{2^{5/2} \sqrt{6}} \, \frac{\Gamma(\alpha+5/2)}{\Gamma(2\alpha)} \sqrt{\frac{\Gamma(2\alpha+n)} {(n+\alpha)\, (n!)^3}} \, \left( \kappa_n q_0 \right)^{\alpha+1/2}
\\
\times  \left(\frac{4}{2\kappa_{n} q_{0} + 3 - 2iq_{0} p_{0}} \right)^{\alpha + \frac{5}{2}} {}_2F_1\left(-n \,, \alpha + \frac{5}{2} \,; 2\alpha \,; \frac{4 \kappa_{n} q_{0}}{2\kappa_{n} q_{0} + 3 - 2iq_{0} p_{0}} \right) \,,
%\end{split}
\end{align}
where ${}_2F_1(-n,b;c;z)$ is a Gauss hypergeometric polynomial of degree $n$. 
%With the value $\alpha = 3/2$ and for the point $(q_{0}, p_{0}) = (1,0)$, this simplifies to
%\begin{equation}
%\label{cnqp32}
%c_n= \frac{\sqrt{27}}{32{\kappa_n}^{2}} \sqrt{\frac{8 \,n!} {\Gamma(5 + n)}} \, \frac{\Gamma(5+n)}{n!} \left( \frac{4 \kappa_{n}}{3+2\kappa_{n}}\right)^{4} F\left(-n \,, 4\,; 5\,; \frac{4 \kappa_{n}}{3+2\kappa_{n}} \right)\,.
%\end{equation}
Hence, we can calculate the time evolution \eqref{distq0p0} for $\alpha = 3/2$ by choosing as an initial state a specific  $(q_0,p_0)$:
\begin{equation}
\rho_{\phi} (q,p,t)= \dfrac{1}{2\pi}\Big| \sum_{n=0}^{n_{\mathrm{max}}}  \overline{c_n(q,p)} \,c_n(q_0,p_0) e^{-i\frac{E_{n}}{\hbar}t} \Big|^{2} \,.
\end{equation}
We recall that the Hamiltonian $A_H$ involved in \eqref{distq0p0} has a continuous spectrum. Therefore the above expression holds as a good approximation only if the initial state $| q_0, p_0 \rangle$ can be essentially represented as a linear combination of bound states. In fact this condition depends  on the choice of $(q_0, p_0)$. A numerical check based on  norm convergence yields $\lim_{n_{max} \to \infty} 2 \pi \, \rho_{\phi} (q_0,p_0,t=0) \simeq1$.\\
We present in Figure \eqref{wpakfig} an example of this dynamical behavior with  initial state taken at $q_0=4$ and $p_0=0$. As expected, the peak of the probability density evolves over the classical trajectory.

In the above calculations, we used the formulae  \cite{magnus66}
\begin{equation}
\label{lagmag1}
\int_0^{\infty}\ud u\, e^{- u}\, u^{\gamma}\, \left(L^{(\gamma-1)}_n (u)\right)^2= (2n+\gamma) \, n! \, \Gamma(\gamma +n)\, , 
\end{equation}
and
\begin{equation}
\label{lagmag2}
\int_0^{\infty}\ud u\, e^{-s u}\, u^{\gamma}\, L^{(\delta)}_n (u) = \frac{\Gamma(\gamma+1)}{n!}\frac{\Gamma(\delta + n+1)}{\Gamma(\delta+1)}\,s^{-\gamma -1}\,{}_2F_1\left(-n,\gamma +1;\delta +1; \frac{1}{s}\right)\, ,  
\end{equation}
which is valid for $\mathrm{Re}\,s> 0\, , \mathrm{Re}\,\gamma> -1$.

\section{Conclusion}
\label{conclus}

We have presented an integral quantization method for the dynamics on the positive half-line. It is based on the affine symmetry of the corresponding half-plane phase space and the related coherent states. The procedure rests upon the resolution of the identity by these states which can be identified with wavelet families in Signal Analysis. This method of quantization differs from the canonical one, $q \mapsto Q$, $p \mapsto P$, and $f(q,p) \mapsto \mathrm{Sym}[f(Q,P)]$. Indeed, canonical quantization (and its more elaborate versions) is based on the \underline{translational} symmetry of the whole plane viewed as the phase space for  the motion on the \underline{whole} line. This is a crucial point which should be always seriously considered in any quantization procedure. Would have we adopted the canonical quantization for the motion on the half-line, we would have never obtained the repulsive centrifugal potential responsible for the regularization of the singularity at the origin of the half-line and for the smooth bouncing of dynamical processes. This fact, which lies at the heart of our results, is illustrated in the present paper with three  illuminating, although quite elementary, examples, the breathing sphere,  the bouncing charged sphere, and dust sphere in cosmology. More elaborate applications of the method are found in recent works related with quantum cosmology  \cite{berdagama14}. 

Another important issue of affine ACS quantization is a systematic rescaling (renormalization?) of quantities depending on the position $q$. This rescaling can be adjusted at will with an original rescaling in the definition of the ACS or/and with the arbitrariness left to us in the choice of  the fiducial vector. As a matter of fact, experiments  or observations determine the constants appearing in the expression of physical quantities, and if these experiments/observations are worked out within the framework of quantum models, their observed numerical values should be consistently inserted in the quantum model, forgetting the classical one. The challenge is now to detect at the scale of our laboratories, for instance with the   optomechanical model mentioned in the introduction,  the appearance of more or less  smooth bouncing when are involved in the formalism positive physical quantities.

%\begin{acknowledgments}
	
%	We gratefully acknowledge Harvey Gould and Jan Tobochnik, who created an earlier 
%	AJP \LaTeX\ sample article that inspired this one.  This work was supported by the 
%	American Association of Physics Teachers.
	
%\end{acknowledgments}

%% The Appendices part is started with the command \appendix;
%% appendix sections are then done as normal sections
%% \appendix

%% \section{}
%% \label{}

%% If you have bibdatabase file and want bibtex to generate the
%% bibitems, please use
%%
%%  \bibliographystyle{elsarticle-harv} 
%%  \bibliography{<your bibdatabase>}

%% else use the following coding to input the bibitems directly in the
%% TeX file.

 \appendix
 \section{\\ A brief review of group transformations and representations}
\label{Grouprep}

A \emph{transformation} of a set $S$ is a one-to-one mapping of $S$ onto itself. A group $G$ is realized as a transformation group of a set $S$ if to each $g \in G$, there is associated a transformation $s \mapsto g\cdot s$ of $S$ where for any two elements $g_1$ and $g_2$ of $G$ and $s \in  S$, we have $(g_1g_2)\cdot s = g_1\cdot (g_2\cdot s)$. The set $S$ is then called a $G$-space. A transformation group is \emph{transitive} on $S$ if, for each $s_1$ and $s_2$ in $S$, there is a $g \in G$ such that $s_2 = g\cdot s_1$. In that case, the set $S$ is called a \emph{homogeneous} $G$-space.

A (linear) \emph{representation} \index{Representation} of a group $G$ is a continuous function $g \mapsto T(g)$ which takes values
in the group of nonsingular continuous linear transformations of a vector space $\mathcal{V}$, and which satisfies the functional equation 
\begin{equation}
T(g_1g_2) = T(g_1)T(g_2) \quad \mathrm{and} \quad T(e) = I\,,
\end{equation}
where $I$ is the identity operator in $\mathcal{V}$ and $e$ is the identity element of $G$. It follows that $T(g^{-1}) = (T(g))^{-1}$. That is, $T(g)$ is a homomorphism of $G$ into the group of nonsingular continuous linear
transformations of $\mathcal{V}$. 

A representation is \emph{unitary}  if the linear operators $T(g)$ are unitary
with respect to the inner product $\lg \cdot |\cdot \rg$ on $\mathcal{V}$. That is, $\lg T(g)\,  v_1| T(g)\, v_2\rg = \lg v_1| v_2\rg$ for all vectors
$v_1$, $v_2$ in  $\mathcal{V}$.  A representation is \emph{irreducible}  if there is no non-trivial subspace $\mathcal{V}_0 \subset \mathcal{V}$ such
that for all vectors $v_o \in  \mathcal{V}_0$, $T(g)\, v_o$ is in $\mathcal{V}_0$ for all $g \in G$. That is, there is no non trivial
subspace $\mathcal{V}_0$ of $\mathcal{V}$ which is invariant under the operators $T(g)$. An important property of unitary irreducible representations (UIR) of a group is the content of Schur's Lemma \cite{barracz77}:
\beprop{\textbf{[Schur's Lemma]}}
\label{schur}
Let $T$ and $T^{\prime}$ be unitary, irreducible representations of $G$ in $\mathcal{V}$ and $\mathcal{V}^{\prime}$, respectively. If $S$ is a bounded linear map of $\mathcal{V} \rightarrow \mathcal{V}^{\prime}$ such that
\begin{equation}
\label{Schurtrans}
ST_{x} = T^{\prime}_{x}S \quad, \quad \forall x \in G \,,
\end{equation}
then, either $S$ is an isomorphism of the spaces $\mathcal{V}$ and $\mathcal{V}^{\prime}$, i.e., $T \backsimeq T^{\prime}$, or $S=0$. Moreover, if $\mathcal{V}=\mathcal{V}^{\prime}$, then  $S$ is a multiple of the identity, $S=cI$. 
\enprop

\section{\\ Affine coherent states quantization: details}
\label{proofresunit}
\subsubsection*{Resolution of the identity by affine coherent states}
Let us introduce the operator $B$ such as
\begin{equation}
\label{Boperator}
B = \int_{0}^{\infty} \int_{-\infty}^{\infty} |q,p\rg \lg q,p| \frac{\ud q \ud p}{2 \pi} \,.
\end{equation}
Then, for arbitrary functions $\phi_{1},\phi_{2} \in \mathcal{H}$ we have
\begin{equation}
\label{Bapplication}
\lg \phi_{1} |B| \phi_{2} \rg = \frac{1}{2\pi} \int_{0}^{\infty} \ud q \int_{-\infty}^{\infty} \ud p \, \lg \phi_{1}|q,p \rg \lg q,p| \phi_{2} \rg \,.
\end{equation}
Using \eqref{affrep+}, we obtain
\begin{equation} \nonumber
\lg \phi_{1} |B| \phi_{2} \rg = \int_{0}^{\infty}\int_{-\infty}^{\infty} \frac{\ud q \ud p }{2\pi q} \int_{0}^{\infty} \int_{0}^{\infty} \ud x_{1} \ud x_{2} \, e^{\ii p(x_{1} - x_{2})}  \overline{\phi_{1}(x_{1})} \psi \left(\frac{x_{1}}{q}\right)  \phi_{2}(x_{2}) \overline{\psi \left(\frac{x_{2}}{q}\right)} \,.
\end{equation}
Since $\int_{\mathbb{R}} dp \, e^{\ii p (x_{1} - x_{2})} = 2\pi \delta (x_{1} - x_{2})$, the integration over $p$ and then over $x_{2}$ gives
\begin{equation}
\lg \phi_{1} |B| \phi_{2} \rg = \int_{0}^{\infty} \frac{\ud q}{q} \int_{0}^{\infty} \ud x \, \overline{\phi_{1}(x)} \phi_{2}(x) \Big\vert \psi \left(\frac{x}{q}\right) \Big\vert^{2} \,.
\end{equation}
Changing the coordinate $q \mapsto q^{\prime} = x/q$, we have
\begin{equation}
\label{identresol}
\lg \phi_{1} |B| \phi_{2} \rg = \int_{0}^{\infty} \frac{\ud q^{\prime}}{q^{\prime}} |\psi(q^{\prime})|^{2} \lg \phi_{1} | \phi_{2} \rg = c_{-1} \lg \phi_{1} | \phi_{2} \rg \,,
\end{equation}
where $c_{-1}$ is a constant given by equation \eqref{cgamma}. This result is a direct consequence of Schur's Lemma \ref{schur}. Therefore, the operator $B$ is proportional to the identity:
\begin{equation}
\label{BeqId}
B = c_{-1} I \, , \quad c_{\gamma}=c_{\gamma}(\psi):=\int_{0}^{\infty}|\psi(x)|^{2}\,\frac{\mathrm{d}x}{x^{2+\gamma}}\,.
\end{equation}
\subsubsection*{Dilating the fiducial vector}
Let us now explore the possibilities offered by  unitary dilations acting on  the fiducial vector $\psi$ and defining the family
\begin{equation}
\label{psika}
\psi_{\kappa}(x) := (U(\kappa, 0) \psi)(x)= \frac{1}{\sqrt{\kappa}}\psi\left(\frac{x}{\kappa}\right)\, , \quad \kappa>0\,. 
\end{equation}
We easily check (with the notation \eqref{ACSdef})  that 
\begin{equation}
\label{acskap}
|q,p\rg_{\psi_{\kappa}}= |\kappa q, p\rg_{\psi}\, , 
\end{equation}
\begin{equation}
\label{cckap}
c_{\gamma}(\psi_{\kappa}) = \frac{1}{\kappa^{2+\gamma}}c_{\gamma}(\psi) \equiv c^{(\kappa)}_{\gamma}\,.  
\end{equation}
Let us consider the quantization map based upon the resolution of the identity obeyed by the ACS $|q,p\rg_{\psi_{\kappa}}$,
\begin{equation}
\label{quantkap}
f(q,p) \mapsto A_{f}^{(\kappa)} = \int_{0}^{\infty} \int_{-\infty}^{\infty} f(q,p) |q,p \rg_{\psi_{\kappa}} {}_{\psi_{\kappa}}\lg  q,p|\, \frac{\ud q\, \ud p}{2\pi c_{-1}(\psi_{\kappa})}  \,.
\end{equation}
The  change of variable $\kappa q \mapsto q$ yields the relation between $A_{f}^{(\kappa)}$ and  $A_f = A_{f}^{(1)}$
\begin{equation}
\label{AfAfkap}
A_{f}^{(\kappa)}= A_{f_{(\kappa)}}\, , \quad f_{(\kappa)}(q,p) := f\left(\frac{q}{\kappa},p\right)\, . 
\end{equation}
Note that this ``dilation covariance" is different of the covariance property \eqref{covaff}.
This gives us an extra degree of freedom besides the choice of the fiducial vector $|\psi\rg$. 
\subsubsection*{$A_{f}^{(\kappa)}$ as an integral operator}
Given two elements $\phi_{1}, \phi_{2} \in \mathcal{H}$, let us determine the corresponding transition matrix element $\lg \phi_{1}| A_{f}^{(\kappa)} | \phi_{2} \rg$ of $A_{f}^{(\kappa)}$. We obtain from  the change of variable $q \mapsto x_1/q$ 
\begin{equation}
\label{matelA}
\lg \phi_{1}| A_{f}^{(\kappa)} | \phi_{2} \rg = \int_{0}^{\infty} \int_{0}^{\infty} \ud x_{1} \ud x_{2} \, \overline{\phi_{1}(x_{1})} \, \mathcal{A}_{f}^{(k)}(x_{1}, x_{2}) \, \phi_{2}(x_{2}) \,,
\end{equation}
where
\begin{equation}
\label{quantaffdilx}
\mathcal{A}_{f}^{(\kappa)}(x_{1}, x_{2}) = \frac{1}{\sqrt{2\pi}c_{-1}(\psi)} \int_{0}^{\infty} \frac{\ud q}{q} \, F_{p} \left(\frac{x_{1}}{\kappa q}, x_{2} - x_{1} \right)\,\psi(q) \, \overline{\psi \left( \frac{x_{2} q}{x_{1}}\right)} \,,
\end{equation}
where $F_p$ stands for the partial inverse Fourier transform
\begin{equation}
\label{parcoure} F_p(q,x)=
\frac{1}{\sqrt{2\pi}}\int_{-\infty}^{+\infty} \ud p e^{-\ii px} f(q,p)\, .
\end{equation}
Hence, the above \eqref{matelA} allows to view $A_{f}^{(\kappa)}$ as the integral operator in $L^2(\R_+\ast,\ud x)$
\begin{equation}
\label{Afintop}
\left(A_{f}^{(\kappa)}\phi\right) (x) = \int_0^{\infty}\ud x^{\prime}\, \mathcal{A}_{f}^{(\kappa)}(x, x^{\prime})\, \phi(x^{\prime})\, , 
\end{equation}
with integral kernel $ \mathcal{A}_{f}^{(\kappa)}(x, x^{\prime})$ given in \eqref{quantaffdilx}. 

For example, if we have a function that depends only on $q$, $f(q,p) = u(q)$, the partial Fourier transform of $f_{(\kappa)}(q,p)$ is
\begin{equation} \nonumber
F_{p}\left( \frac{x_{1}}{\kappa q}, x_{2} - x_{1} \right) = \sqrt{2 \pi} \, \delta(x_{2} - x_{1} ) \, u \left( \frac{x_{1}}{\kappa q} \right) \,.
\end{equation}
Then the kernel \eqref{quantaffdilx} reads as
\begin{equation}
\label{quantaffkappaq}
\mathcal{A}_{u}^{(\kappa)}(x_{1}, x_{2}) = \frac{1}{c_{-1}} \delta(x_{2} - x_{1}) (|\psi|^{2} \ast_{\mathrm{aff}} u) \left( \frac{x_{1}}{\kappa } \right) \,,
\end{equation}
where we have introduced the (commutative) convolution $\ast_{aff}$,
\begin{equation}
\label{convdef}
(f_{1} \ast_{aff} f_{2}) (x) = \int_{0}^{\infty} \frac{\ud q}{q} f_{1}(q) f_{2} \left( \frac{x}{q}\right) = (f_{2} \ast_{aff} f_{1}) (x)\,.
\end{equation} 
\eqref{quantaffkappaq} simply means that $A_{u}^{(\kappa)}$ is the multiplication operator
\begin{equation}
\label{Aukap}
A_{u}^{(\kappa)} = (|\psi|^{2} \ast_{\mathrm{aff}} u) \left( \frac{Q}{\kappa } \right)\, , 
\end{equation}
and $Q$ is the multiplication operator, $Q\phi(x) = x \phi(x)$. 
For the important case $u(q)=q^{\beta}$, the operator $A_{q^{\beta}}^{\kappa}$ assumes the simple expression 
\begin{equation}
\label{Akappaqpowers}
A_{q^{\beta}}^{(\kappa)} = \frac{c_{\beta - 1}(\psi)}{c_{-1}(\psi)} \frac{Q^{\beta}}{\kappa^{\beta}} \,,
\end{equation}
The other important particular  case holds when the function $f$ depends on $p$ only, $f(q,p)= v(p)$. Then the integral kernel is independent of $\kappa$ and becomes
\begin{equation}
\label{quantaffkappap}
\mathcal{A}_{v}^{(\kappa)}(x_{1}, x_{2}) = \mathcal{A}_{v}(x_{1}, x_{2})= \frac{1}{\sqrt{2\pi}c_{-1}(\psi)} \hat v(x_{2} - x_{1}) (\psi \ast_{\mathrm{aff}} \widetilde{\bar \psi}) \left( \frac{x_{1}}{x_2} \right) \,,
\end{equation}
where $\hat v(x) = \frac{1}{\sqrt{2\pi}}\int_{-\infty}^{+\infty}e^{-\ii px}\, v(p)\,\ud p$ is the Fourier transform of $v$ and $\widetilde{\psi}(x):=\psi(1/x)$. Hence, to $v(p) = p^n$, $n\in \N$,  corresponds the operator
\begin{equation}
\label{Akappappowers}
A_{p^n}^{(\kappa)}= A_{p^n}= \frac{1}{c_{-1}(\psi)}\sum_{m=0}^n \binom{n}{m} c_{m-n-1}^{(n-m)}\, \frac{(-\ii)^{n-m}}{Q^{n-m}}\, P^m\, , \quad P= -\ii \frac{\partial}{\partial x}\, , 
\end{equation}
where we have introduced the convenient notation which extends \eqref{BeqId},
\begin{equation}
\label{cgamm}
c_{\gamma}^{(m)}(\psi):= \int_0^{\infty} \frac{\ud x}{x^{\gamma + 2}}\, \psi(x)\,\overline{\psi^{(m)}}(x)\,, \quad  c_{\gamma}^{(0)}(\psi)= c_{\gamma}(\psi)\, .
\end{equation}
Applied to the lowest powers (relevant to this paper), \eqref{cgamm} yields the expressions
\begin{equation}
\label{AApp2}
A_p= P-\frac{\,c_{-2}^{(1)}(\psi)}{c_{-1}(\psi)}\frac{\ii}{Q}\, , \quad A_{p^2} = P^2 - \frac{\,c_{-2}^{(1)}(\psi)}{c_{-1}(\psi)}\frac{2\ii}{Q}\, P - \frac{c_{-2}^{(1)}(\psi)}{c_{-1}(\psi)}\frac{1}{Q^2}\, , 
\end{equation}
When $\psi$  is real, we have (from integration by part and boundary values) $c_{-2}^{(1)}(\psi)= 0$ and $c_{-3}^{(2)}(\psi)= - \int_0^{\infty} \ud x\, x\, \left(\psi^{\prime}(x)\right)^2$. Hence, \eqref{AApp2} reduces to
\begin{equation}
\label{AApp2real}
A_p= P\, , \quad A_{p^2} = P^2 +\frac{K_{\psi}}{Q^2}\, , \quad  K_{\psi}= \frac{1}{c_{-1}(\psi)}\int_0^{\infty} \ud x\, x\, \left(\psi^{\prime}(x)\right)^2 >0\, . 
\end{equation}

\subsubsection*{Lower symbols}

We now give details about the calculation of lower symbols introduced in \eqref{expecAf}
\begin{equation}
\label{expecAfA}
\check f (q,p) = \lg q,p |A_f| q,p\rg \,.
\end{equation}
 Supposing that inverting the order of the integrals is legitimate here, we obtain
\begin{align}
\nn
\check{f}(q,p)= \frac{1}{\sqrt{2\pi}c_{-1}(\psi)} \int_{0}^{\infty} \frac{\ud q^{\prime}}{qq^{\prime}} \int_{0}^{\infty} \int_{0}^{\infty} \ud x\, \ud x^{\prime}  \left[ e^{\ii p(x-x^{\prime})}
\,F_p(q^{\prime},x-x^{\prime}) \cdot \right.
\\
\label{afflowsymb}
\left. \psi\left(\frac{x}{q}\right)\, \overline{\psi}\left(\frac{x}{q^{\prime}}\right)
\psi\left(\frac{x^{\prime}}{q^{\prime}}\right)\,\,
\overline{\psi}\left(\frac{x^{\prime}}{q}\right) \right]\,,
\end{align}
where $F_p$ is given by \eqref{parcoure}.  

For a function  depending  on $q$ only, $f(q,p) = u(q)$, this integral is does not depend on $p$ and is expressed in terms of the inner product in $L^2(\R_+^\ast,\ud x)$ and the affine convolution as
\begin{equation}
\label{czechu}
\check{u}(q)= \frac{1}{c_{-1}(\psi)}\left\lg \frac{1}{q}\left\vert \psi\left(\frac{\cdot}{q}\right)\right\vert^2 \, \big| \, u \ast_{\mathrm{aff}}\vert\psi\vert^2 \right\rg\, . 
\end{equation}
Applied to powers of $q$, this formula simplifies to
\begin{equation}
\label{czechqbeta}
\check{q^\beta}= \frac{c_{\beta-1}(\psi)\,c_{-\beta-2}(\psi)}{c_{-1}(\psi)}\,q^\beta\, . 
\end{equation}
And, applied to  nonnegative integer powers of $p$, \eqref{afflowsymb} leads to the polynomial expansion in powers of $p$ with coefficients which are proportional to inverse powers of $q$:
\begin{equation}
\label{czechpn}
\check{p^n}=  
\frac{1}{c_{-1}(\psi)}\sum_{0\leq m+m^{\prime}\leq n}\frac{n!}{m!m^{\prime}!(n-m-m^{\prime})!}\, c_{-m^{\prime}-1}^{(m^{\prime})}(\psi)\,\overline{c_{m^{\prime}-2}^{(m)}}(\psi)\, (-\ii)^{m+m^{\prime}}\,\frac{p^{n-m-m^{\prime}}}{q^{m+m^{\prime}}} \, , 
\end{equation}
where the constants $c_{\gamma}^{(m)}(\psi)$ were introduced in \eqref{cgamm}. For $n=1$ and $n=2$, and with real $\psi$, this formula simplifies to
\begin{eqnarray}
\label{pp2}
\check{p} &=& p\, ;
\\
\check{p^2} &=& p^2 - \left(c_{-2}^{(2)}(\psi) + \frac{c_{-3}^{(2)}(\psi)c_{0}(\psi)}{c_{-1}(\psi)}\right)\frac{1}{q^2}= p^2+ c(\psi)\frac{1}{q^2}\, , 
\end{eqnarray}
where $c(\psi)$ was defined  in \eqref{symbp2}. 

\section{\\ Covariant integral quantizations} 
\label{covintquant}

Lie group representations \cite{barracz77} offers a wide range of possibilities for implementing integral quantization(s).  Let $G$ be a Lie group with left Haar measure $\mathrm{d}\mu(g)$, i.e. $\mathrm{d}\mu(g_0g)= \mathrm{d}\mu(g)$ for all $g_0 \in G$, and let $g\mapsto U\left(g\right)$ be a unitary irreducible representation (UIR) of $G$ in a Hilbert space $\mathcal{H}$. Consider a bounded
operator $\mathsf{M}$ on $\mathcal{H}$ and suppose that the operator
\begin{equation}
\label{boundR1}
R:=\int_{G}\mathsf{M}\left(g\right)\mathrm{d}\mu\left(g\right),\
\mathsf{M}\left(g\right):=U\left(g\right)\mathsf{M}U^{\dagger}\left(g\right)\,,
\end{equation}
is defined in a weak sense, i.e. 
\begin{equation}
\label{boundR2}
\lg \phi_1| R \phi_2\rg =\int_{G}\lg \phi_1|\mathsf{M} \phi_2\rg \left(g\right)\mathrm{d}\mu\left(g\right)\,,
\end{equation}
for all $\phi_1, \phi_2$ in a dense subset of $\mathcal{H}$. From the left invariance of $\mathrm{d}\mu(g)$ we have
\begin{equation}
\label{comRU}
U\left(g_{0}\right)R\,U^{\dagger}\left(g_{0}\right)=\int_{G}\mathsf{M}\left(g_{0}g\right) \mathrm{d}\mu\left(g\right)=R\,,
\end{equation}
so $R$ commutes with all operators $U(g)$, $g\in G$. Thus, from Schur's Lemma in the \eqref{schur}, $R=c_{M}I$ with
\begin{equation}
\label{cM}
c_{M}=\int_{G}\mathrm{tr}\left(\rho_{0}\mathsf{M}\left(g\right)\right)\mathrm{d}\mu\left(g\right)\,,
\end{equation}
where the unit trace positive operator $\rho_{0}$ is chosen in order to make the integral converge. This family of operators provides the resolution of the identity on $\mathcal{H}$.
\begin{equation}
\label{eq:resolution}
\int_{G}\mathsf{M}\left(g\right)\mathrm{d}\nu\left(g\right)=I,\qquad
\mathrm{d}\nu\left(g\right):=\frac{\mathrm{d}\mu\left(g\right)}{c_{M}}
\end{equation}
and the subsequent quantization of complex-valued functions (or distributions, if well-defined) on $G$
\begin{equation}
\label{quantgr} 
f\mapsto A_{f}=\int_{G}\, \mathsf{M}(g)\, f(g)\,\mathrm{d}\nu(g)\,.
\end{equation}
This linear map, function $\mapsto$ operator in $\mathcal{H}$, is covariant in the sense that
\begin{equation}
\label{covarG} 
U(g)A_{f}U^{\dagger}(g)=A_{\mathfrak{U}(g)f}\,.
\end{equation}
where $(\mathfrak{U}(g)f)(g^{\prime}) :=f(g^{-1}g^{\prime})$. In the case when $f\in L^{2}(G,\mathrm{d}\mu(g))$, the latter is the regular unitary representation.

A semi-classical analysis \cite{klauder12,klauderbook15} of the operator $A_f$ can be implemented through the study of new functions, denoted by $\check f$, on $X$. They are a generalisation of objects called lower symbols by Lieb \cite{lieb73} and covariant symbols by Berezin \cite{berezin74}.  Suppose that $\mathsf{M}$ is a density, i.e. non-negative unit-trace, operator $\mathsf{M}= \rho$ on $\mathcal{H}$. Then the operators $\rho(g)$ are also densities, and this allows to build the function $\check{f}(g)$  as
\begin{equation}
\label{lowsymb} 
\check{f}(g)\equiv \check{A_f}:=\int_{G}\,\mathrm{tr}(\rho(g)\,\rho(g^{\prime}))\,
f(g^{\prime})\mathrm{d}\nu(g^{\prime})\,.
\end{equation}
The map $f\mapsto \check{f}$ is a generalization of the Berezin or
heat kernel transform on $G$ \cite{hall06}.  

Let us illustrate the above procedure with the case of square integrable UIR's and rank one $\rho$. For a square-integrable UIR $U$ for which $\left\vert \psi\right\rangle $ is an admissible
unit vector, i.e.,
\begin{equation}
\label{cpsi} 
c(\psi):=\int_{G}\mathrm{d}\mu(g)\,|\left\langle \psi\right\vert U\left(g\right)\left\vert \psi\right\rangle |^{2}<\infty\,,
\end{equation}
the resolution of the identity is obeyed by the coherent states
$\left\vert \psi_{g}\right\rangle =U(g)\left\vert \psi\right\rangle$, in a generalized sense, for the group $G$:
\begin{equation}
\label{residsq} 
\int_{G}\rho(g)  \mathrm{d}\nu\left(g\right)=I\,\quad\mathrm{d}\nu\left(g\right)=
\frac{\mathrm{d}\mu\left(g\right)}{c(\psi)}\, , \quad\rho(g)=\left\vert \psi_g\right\rangle \left\langle\psi_g\right\vert \,.
\end{equation}
Choosing as $\mathsf{M}$ a density operator $\rho$, as we did in this case, has multiple advantages,  peculiarly in  regard to probabilistic aspects both on classical and quantum levels \cite{gazhell15}.

\section*{Acknowledgments}
J.-P. Gazeau thanks CBPF and CNPq (Brazil) for financial support and CBPF for hospitality. C.R. Almeida thanks CAPES (Brazil) and A.C. Scardua also thanks CNPq for partial financial support.

\end{document}